\documentclass[fleqn,usenatbib]{mnras}

\usepackage{newtxtext,newtxmath}

\usepackage[T1]{fontenc}
\usepackage{ae,aecompl}

\usepackage{graphicx}	
\usepackage{amsmath}	

\title[Simulations of Globular Clusters]{Simulations of Globular Clusters Within Their Parent Galaxies: Multiple Stellar Populations And Internal Kinematics}

\author[McKenzie \& Bekki]{
Madeleine McKenzie,$^{1}$\thanks{E-mail: madeleine.mckenzie@icrar.org}
Kenji Bekki,$^{1}$
\\
$^{1}$ ICRAR, M468, The University of Western Australia, 35 Stirling Highway, Crawley, WA 6009, Australia
}

\date{Accepted XXX. Received YYY; in original form ZZZ}

\pubyear{2020}

\begin{document}
\label{firstpage}
\pagerange{\pageref{firstpage}--\pageref{lastpage}}
\maketitle

\begin{abstract}
Using three-dimensional smoothed particle hydrodynamics simulations, we investigate the formation of multiple stellar populations (MSPs) in globular clusters (GCs) within the context of their parent galaxies. In our scenario, the second generation (2G) of stars originate from both asymptotic giant branch (AGB) polluters and pristine gas accreted from the host galaxy. Previous theoretical and numerical studies have demonstrated that this ``AGB with dilution'' model has the potential to alleviate several problems faced by the classical AGB scenario. However, the accretion of pristine gas onto the GC has yet to be investigated within the context of the parent galaxy. This paper presents the preliminary results from our original simulation code which models GC formation from giant molecular clouds in a host galaxy, and subsequent gas accretion onto the GC. By simulating the genesis of the 2G over a 370 Myr time frame, we demonstrate that the fraction of 2G stars is inextricably linked to the GC's environment. Our simulations rationalise the wide variety of abundance patterns, kinematics and 2G concentrations by altering the initial conditions of both the GC progenitor and the host galaxy itself. Most notably, we reproduce a positive correlation between the fraction of 2G stars and the initial mass of the cluster. We discuss the physical implications of our scenario and compare our simulations with observations of the Galactic GC 47 Tucanae (47 Tuc). Finally, we present scaling relations which encompass the wider GC population and serve as a reference for future observations.

\end{abstract}




\section{Introduction}

The answer to how multiple stellar populations (MSPs) form within globular clusters has evaded astronomers for several decades. Since the suggestion that MSP may be responsible for the Cyanogen bimodality in 47 Tucanae (47 Tuc) by \cite{Norris&freeman1979}, a wealth of photometric (e.g. \citealt{Sbordone_etal2011}; \citealt{Piotto_etal2015}; \citealt{Milone_etal2017}; \citealt{Lagioia_etal2019}; \citealt{Lee_Sneden_2020}) and spectroscopic (e.g. \citealt{Barbuy_etal2009}; \citealt{Carretta_etal2009}; \citealt{Kamann_etal2018}; \citealt{Latour_etal2018}; \citealt{Husser_etal2020}) evidence indicates that MSPs are the source of light element inhomogenities almost ubiquitously observed in GCs. Several groups and individuals have postulated various scenarios which fulfil observational criteria to varying degrees, but none have been universally accepted. 

The majority of observed GCs exhibit anti-correlations between Na and O (e.g. \citealt{Carretta_2019}) and occasionally Al and Mg (\citealt{Pancino_2017}). Subtle differences in light element abundances are used to delineate different populations of stars. Enriched populations are characterised by enhanced N, Na and Al and depleted C, O and Mg. The terminology for such population varies widely in the literature, with first and second generations (1G and 2G respectively, introduced in \cite{Piotto_etal2015}) commonly being employed. Such nomenclature assumes a linear formation scenario which has yet to be confirmed, however, we adopt it for continuity. 1G stars show primordial (oxygen-rich, sodium-poor) chemical composition similar to that of field stars (e.g. \citealt{Martell_2011}). Elemental abundance patterns of 2G stars are ascribed by CNO cycling and p-capture processes at high temperatures (e.g. \citealt{Gratton_etal2012}). It has been hypothesised that the chemical abundance patterns of 2G stars are specific to GCs, however, N-enriched stars akin to the 2G have been identified by the APOGEE survey within the galactic bulge \citep{Schiavon_etal2016}. One explanation for this is that these 2G stars are the remains of tidal interactions between GCs and the Milky Way (\citealt{Vesperini_etal2010}; \citealt{Martell_etal2016}).

Further classifications exist for the general GC population. Type I GCs constitute $\approx83\%$ of all clusters \citep{Marino_etal2019} and show a narrow spread in iron (e.g. \citealt{Carretta_2009_Fe_scatter}; \citealt{Bailin_2019}). Type II GCs show evidence of several groups of stars with varying iron abundances with notable examples including $\omega$ Centauri (\citealt{Lee_etal1999}; \citealt{Bedin_etal2004}), M54 (\citealt{Carretta_etal2010}; \citealt{Layden_etal2000}) and NGC1851 \citep{Carretta_etal2011}. Popularised by \cite{Milone_etal2017}, pseudo two-color diagrams known as `chromosome maps' have become a powerful tool for diagnosing and quantifying multiple populations, for example, four populations have been identified in NGC2080 (\citealt{Milone_etal2015_2808}; \citealt{Latour_etal2019}) and five in M15 \citep{Nardiello_etal2018} using this technique. By exploiting small chemical differences between stars, it is clear that the notion of single stellar populations must be discarded.

Due to the wide variety of physical and dynamical conditions observed in GCs, a robust yet flexible scenario is required in order to encapsulate their formation processes. \cite{Milone2019_13facts} concisely summarises the observational criteria for which a scenario must explain. For example, the scenario must be capable of reproducing a central concentration of 2G stars (e.g. 47 Tuc in  \citealt{Milone_etal201847Tuc} and NGC3201 in \citealt{Carretta_etal2010NGC3201}), equivalent concentrations of 1G and 2G stars (e.g. M71 in \citealt{Gerber_etal2020}) and on rarer occasions, a centrally concentrated 1G (i.e. M15 in \citealt{Larson_etal2015} or M80 in  \citealt{Dalessandro_etal2018M80}). It should reproduce a fraction of 2G stars which varies from as low as 25\% (NGC339 in \cite{Niederhofer_etal2017}) to 90\% in $\omega$ Centauri \citep{Bellini_etal2009}. Additionally, there should be a way of explaining the stellar halo around M2 \citep{Kuzma_etal2016_m2} and NGC1851 \citep{Kuzma_etal2018}.

Several reviews by \cite{Gratton_etal2012}, \cite{Renzini_etal2015}, \cite{Bastian&lardo2018} and most recently \cite{Gratton_etal2019}
summarise the leading formation scenarios for GCs. Many of the canonical formation scenarios have been plagued by a variety of unexplained observational patterns. For example, the AGB scenario first discussed by \cite{Iben_1975}, \cite{Iben_1976} and later in \cite{Cottrell_DaCosta_1981} succeeds in justifying a number of observations (e.g. \citealt{Ventura_DAntona2008}; \citealt{Ventura_DAntona2008Metal}; \citealt{Choi_Yi2008}; \citealt{DOrazi_etal2010}; \citealt{DErcole_etal2012}; \citealt{DAntona_etal2016}) but suffers from two main criticisms. The "mass budget problem" (e.g. \citealt{Smith_norris_1982}, \citealt{Renzini2008}) is where the progenitor mass of the 1G is not large enough to form a massive 2G. This issue could be alleviated by employing a top-heavy IMF (\citealt{Prantzos_Charbonnel2006}; \citealt{Bekki_etal2017}; \citealt{Kroupa_2019}) or through assuming a mechanism by which a significant number of primordial stars are stripped (e.g. \citealt{Decressin_etal2008}). Secondly, AGB models predict a Na-O correlation rather than anti-correlation \citep{Bastian&lardo2018}. A solution to these problems, suggested by \cite{DErcole_etal2008}, was to allow AGB ejecta to mix with the pristine gas with comparable metallicities to the 1G. Additional mass is obtained through accretion of gas from the parent galaxy and the enrichment patterns of 2G stars are thus dependent on the degree of mass accretion as well as AGB yields. Several other polluters besides AGB stars have been accused of producing MSPs; including fast-rotating massive main sequence stars (\citealt{Decressin_etal2007}), binary mergers (\citealt{deMink2009}), supermassive main sequence stars (\citealt{Denissenkov_Hartwick2014}) and cool supergiant stars (\citealt{Szecsi_Wunsch2019}).

Recently, the scenario proposed by \cite{wang_etal2020} assumes a multichannel scenario where not one particular polluter is responsible for the formation of enriched populations. Through dynamical modelling of binary mergers, supermassive and fast-rotating massive main sequence stars, they generate a chemically enriched population analogous to observed Galactic GCs. As there is no need for gas accretion, they state that their scenario is independent of the galactic conditions of their parent galaxy. However, their scenario is at odds with recent results from \cite{milone_etal2020} where they found that Magellanic Cloud GCs had a notably lower 2G fraction than what was predicted from the mass to enriched fraction found in Galactic GCs.

The hierarchical formation model of $\Lambda$CDM dictates that our Galaxy underwent several merging events in its history (e.g. \citealt{Belokurov_etal2018}; \citealt{Helmi_etal2018}; \citealt{Myeong_etal2019}; \citealt{Kruijssen_etal2020}). Merging episodes brought a considerable number of GCs into our galaxy (e.g. \citealt{Myeong_etal2018SausageGC}; \citealt{Horta_etal2020}; \citealt{Forbes_2020}) several of which have since been traced back to these events. These GCs have been identified through velocity and action space calculations, and APOGEE surveys have chemically linked these clusters as well. If the fraction of enriched stars formed within GCs is mediated by the host galaxy as found by \cite{milone_etal2020}, a formation scenario should be capable of accounting for these differences. 

Recent hydrodynamical simulations of GC formation have demonstrated that feedback effects from SNe and AGB stars within fractal giant molecular clouds (GMCs) influence the chemical and dynamical properties of 2G stars (\citealt{Bekki_2017}; \citealt{Bekki_2017Cloud}). \cite{Calura_etal2019} recently investigated the mixing process of accreting interstellar medium (ISM) and AGB ejecta in a self-consistent manner and reproduced the observed He abundance spread in GCs. Using new hydrodynamical simulations with a mass resolution of $\sim 0.01 {\rm M}_{\odot}$, \cite{Bekki_2019} investigated the impact of direct interaction of individual 1G stars with intra-cluster gas. Due to computational complexities, simulations of MSP formation tend to neglect the impact of the parent galaxies dynamics. This oversimplification hinders the investigation of a number of key observations such as the correlation between GC masses and 2G fraction (\citealt{Milone_etal2017} and references therein) and internal rotation of the 1G and 2G populations (e.g. \citealt{Bianchini_etal2013} \citealt{Bellini_etal2015}; \citealt{Bianchini_etal2018}; \citealt{Bellini_etal2018}, \citealt{Cordoni_etal2020}).

The purpose of this paper is to discuss our preliminary findings from our original simulation code. We present a simulated analogue of the Galactic GC 47 Tuc, an archetypal Type I GC, to illustrate the success and drawbacks of this model. We trace the evolution of our fiducial 47 Tuc analogue over a 370 Myr window and analyse the morphological, chemical and kinematic consequences of gas accretion onto the cluster. Next, we extend our analysis to a broader population of GC and parent galaxy combinations and demonstrate that the gas fraction and density of the galaxy mediates the relationship between the fraction of 2G stars and total mass of the cluster as found by \cite{milone_etal2020}. Given the presence of rotation in a number of clusters (e.g. see \citealt{sollima_etal2019}) we describe the correlations between the 1G and 2G kinematics and the total cluster mass. Finally, we discuss how these simulations can naturally give rise to multiple discrete 2G populations and Type II GCs. 

The simulation code employed in the present study allows us to investigate how complex stellar clusters can form within their natal GMCs in dwarf galaxies. Previous theoretical and numerical studies have investigated the physical roles of external gas accretion onto GCs in detail (e.g., \citealt{Bekki_Mackey2009}; \citealt{Pflamm-Altenburg_Kroupa2009}; \citealt{Conroy_Spergel2011}; \citealt{Armstrong_etal2018}; \citealt{McKenzie_Bekki2018}). However, these works did not self-consistently investigate the formation of 1G stars and subsequent gas accretion on the existing stellar systems. Thus it is unclear how additional populations of stars can form and how the properties of the parent galaxies impact such populations. Over a series of papers, we intend to develop our understanding of GC formation and trace the impact of the parent galaxy throughout the cluster's evolution. 

We structure the paper as follows. In Section \ref{model}, we describe our original simulation code. In Section \ref{results}, we present a 47 Tuc analogue for our scenario and global scaling relations from our scenario. In Section \ref{discussion}, we discuss the ramifications of our simulations and summarise our conclusions in Section \ref{conclusion}.

\section{The model}
\label{model}
\subsection{A GC formation  scenario}

In the standard AGB scenario, an essential physical process of GC formation is the efficient conversion of ISM mixed with AGB ejecta into new stars. In the present study, we assume these mixing processes and subsequent star formation can occur within high redshift gas-rich dwarf galaxies. We use galaxy-scale hydrodynamical simulations to investigate the formation of 1G and 2G stars simultaneously; however, some physical processes are poorly understood (e.g.,  dynamical and thermal influences of multiple SNe on GC-hosting GMCs). Accordingly, we adopt a somewhat idealized model for various physical processes related to GC formation. We describe these simplifications and discuss possible shortcomings of our GC formation model in Section \ref{discussion} and suggest ways of constructing more realistic models in future works.

\subsubsection{Formation of massive molecular clouds in dwarfs}

In order to explain the observed typical mass ($\sim 3 \times 10^5 {\rm M}_{\odot}$) and high fraction of 2G stars ($\sim 0.7$), the original mass of GCs has been suggested to be $1-3 \times 10^6 {\rm M}_{\odot}$ (the so-called mass budget problem). This implies that natal GMCs should be at least $[5-15] \times 10^6 {\rm M}_{\odot}$ when assuming a star formation efficiency within the GMCs of around 20\%. Therefore, only gas-rich dwarf galaxies with the potential to form massive GMCs can host GCs in the present formation scenario. We attribute the formation of massive GMCs to gravitational instabilities in the stellar and gaseous discs of dwarf galaxies with high baryonic fractions $f_{\rm b}$.
Hierarchical merging of numerous smaller GMCs is required to generate a cloud more massive than $10^6 {\rm M}_{\odot}$ which may eventuate into a GC. Merging or collisions of two GMCs may 
trigger the formation of massive OB stars in nearby galaxies
(e.g., \citealt{Fukui_etal2017}; \citealt{Tsuge_etal2019}).
Hence, multiple merging or collision events could produce a large number of OB stars which may influence the evolution of GC-hosting GMCs through energetic feedback effects.

\subsubsection{Formation of 1G stellar systems and the subsequent
dispersion of GMCs by 1G SNe}

Following the creation of massive GMCs, 1G stars within a cloud can efficiently form within short time-scales of a few Myrs. However, the formation process leaves behind a significant fraction of the original gas. 1G stars quickly develop into compact stellar systems, although the dynamical relaxation process is still incomplete after 10 Myr. The most massive of these stars ($m \sim [50-100] {\rm M}_{\odot}$) explode as Type II SNe in $\sim 3$ Myr after the formation of 1G stars. A large amount of thermal and kinetic energy from these multiple SNe can disperse the remaining gas completely, and the SN feedback effects can form a giant gaseous hole in its host dwarf galaxy. Such giant HI holes are observed in some dwarf galaxies, and this is discussed further in Section  \ref{sec:Large_HI}.

\subsubsection{Accretion of ISM and AGB ejecta onto 1G systems}

After the complete dispersal of GC-hosting GMCs, 1G stellar systems
can gravitationally attract their surrounding ISM and provided that they are massive enough, can accrete a significant portion of gas. The stellar winds of massive AGB stars can also be gravitationally trapped by the stellar systems and can mix with the accreted ISM. The accretion process of ISM can proceed sporadically, and it is not like ``Bondi-type'' \citep{Bondi_1952} continuous accretion as the GC-host dwarfs contain clumpy ISM consisting of numerous small and large gas clouds. This clumpy accretion can result in discrete multiple stellar populations, as discussed later.

\subsubsection{2G formation within 1G systems}

Once the 1G accumulates some critical mass of ISM and AGB ejecta in its central region, new stars can be formed efficiently from the gas within the gravitational potentials of the systems. These 2G stars can have more compact spatial distributions compared to their 1G counterparts, as demonstrated in previous numerical simulations of GC formation with MSPs (e.g., \citealt{Bekki_2011}). However, it could be possible that 2G stars can have less compact distributions than the 1G provided that the accreting ISM has a considerable amount of angular momentum with respect to the centre of 1G stellar systems. The mass-ratios of 2G to 1G stars can possibly depend on the ISM accretion processes and thus on the physical properties of GC-hosting dwarfs.

\subsection{Simulation code}

In order to investigate the details of the above key formation
processes of GCs with MSPs, we use our original chemodynamical simulation code based on the SPH method that can run on GPU machines (\citealt{Bekki_2013}; \citealt{Bekki_2015}). This code allows us to investigate the formation of dust from SNe and AGB stars, the destruction of dust by SNe in ISM, the formation of molecular hydrogen on dust grains,  chemical evolution of various elements, star formation (on galaxy-scale), and various feedback effects from stars. However, simulating these complicated processes of ISM is numerically costly. Because the main focus of this study is not the formation and evolution of dust and molecular hydrogen, we switch off calculations related to dust and molecular hydrogen formation. We also newly introduce our original multiple gravitational lengths with very narrow maximum time step width to investigate gas accretion processes on GCs as described below.

\subsection{Gas-rich dwarf galaxy}

A GC-forming  dwarf galaxy is assumed to consist of a
dark matter halo (DM), stellar disc,  and gaseous disc.
The initial total masses of DM, stellar disc, and gas disc
are denoted as $M_{\rm dm}$, $M_{\rm s}$, and $M_{\rm g}$,
respectively. We mainly investigate low-mass dwarf disc galaxies in which
(i) $M_{\rm dm} = [1-5] \times 10^{10} {\rm M}_{\odot}$ and
(ii) the mass ratio of the disc ($M_{\rm s}+M_{\rm g}$)
to the dark matter halo ($M_{\rm h}$) in a dwarf disc galaxy ranges from 0.01 to 0.06. The adopted small baryonic fractions are quite reasonable, given that recent observations have found that low-mass halos have smaller $f_{\rm b}$.

We adopt the `NFW' profile for the dark matter halo (\citealt{NFW_1996}) with a central cusp predicted by the Cold Dark Matter (CDM)  model:
\begin{equation}
{\rho}(r)=\frac{\rho_{0}}{(r/r_{\rm s})(1+r/r_{\rm s})^2},
\end{equation}
where $r$,  $\rho_{0}$,  and $r_{\rm s}$ are the distance from the centre
of the cluster, the central density, and the scale-length of the dark halo,
respectively. The virial radius ($r_{\rm vir}$),  the scale radius ($r_{\rm s}$), and the `$c$' parameter (=$r_{\rm vir}/r_{\rm s}$) are chosen such that the values are consistent with recent cosmological simulations
for the adopted $M_{\rm h}$ (\citealt{Neto_etal2007}). We mainly investigate the models with $c=16$, which is reasonable for low-mass halos.

The mass and size of the galactic bulge in a disc galaxy
are free parameters denoted as $M_{\rm b}$ and $R_{\rm b}$, respectively. The radial ($R$) and vertical ($Z$) density profiles of the
adopted exponential stellar disc are assumed to be proportional to $\exp (-R/R_{0}) $ with scale length $R_{0} = 0.2R_{\rm s}$  and to ${\rm sech}^2 (Z/Z_{0})$ with scale length $Z_{0} = 0.04R_{\rm s}$, respectively. The gas disc with a size  $R_{\rm g}=R_{\rm s}$
has the radial and vertical scale lengths of $0.2R_{\rm g}$ and $0.02R_{\rm g}$, respectively. The disc of the present model has
$R_{\rm s}=17.5$ kpc and in addition to the rotational velocity caused by the gravitational field of disc, bulge, and dark halo components. The initial radial and azimuthal velocity dispersions are assigned to the disc component according to the epicyclic theory with Toomre's parameter $Q$ = 1.5. The gas mass fraction ($f_{\rm g}=M_{\rm g}/M_{\rm s}$) is also a free parameter in the present study.

We mainly investigate models where the initial gaseous temperature and metallicity in the ISM is set to 100K and [Fe/H]=$-0.8$ respectively,
as we wish to find the most suitable model for 47 Tuc with the observed
metallicity of [Fe/H]$\sim -0.8$. However, we investigate the models with different [Fe/H] in the ISM. The radiative cooling processes are properly included by using the cooling curve by \cite{Rosen_Bregman_1995} for  $T < 10^4$K and the MAPPING III code for $T \ge 10^4$K \citep{Sutherland_Dopita_1993}.
Heating of the ISM by dust may be significant in its evolution within galaxies (e.g., photo-electric heating; \citealt{Osman_etal2020PEH}),
but has been excluded due to the cost incurred by additional calculations in the hydrodynamical simulations.

\subsection{Structure and kinematics of the 1G system}
The 1G stellar system of the GC is assumed to have a mass of $M_{\rm gc}$, a size of $R_{\rm gc}$, and a radial scale-length of $0.2R_{\rm gc}$. The GC is represented by a rotating Plummer model where $M_{\rm gc}$ and $R_{\rm gc}$ are free parameters. The GC is assumed to be supported by velocity dispersion and rotation and the dispersion is assumed to be isotropic. We introduce the following parameter to describes the ratio of total rotational energy ($T_{\rm rot}$) of a GC to its total kinetic energy ($T_{\rm kin}$):
\begin{equation}
f_{\rm rot}= \frac{ T_{\rm rot} }{ T_{\rm kin} } .
\end{equation}
We assume that the GC's spin axis is inclined by $\theta_{\rm gc}$ degrees with respect to the spin axis of its host galaxy (the $z$-axis).

\subsection{Giant HI hole after GC formation}

A very high star formation efficiency within the GMCs is required to allow for the creation of bound star clusters. A small fraction of GC-forming gas clouds may remain within the cluster after 1G formation; however, multiple SNe from the 1G should expel all of the remaining gas. Thus we assume that the natal GMC of the cluster is both consumed by efficient 1G formation and dispersed by energetic SN feedback effects after 1G formation. Our current hydrodynamical simulation can not achieve the required spatial resolution to investigate how multiple SNe events starting from just 3 Myr after the 1Gs creation can influence the remaining gas trapped within the cluster. We therefore approximate this process by artificially inducing a giant gaseous hole after 1G formation centred on the birthplace of a GC. We admit that this is an over-simplification of GC formation, however, we believe that this allows us to investigate how ISM and AGB ejecta can be accreted onto existing 1G systems. The size of the gas hole ($R_{\rm HI}$) is a free parameter ranging from 100 pc to 1 kpc.

\subsection{Formation model for 2G stars}

In our scenario, new stars can form through the accretion of ISM onto the 1G or from gas ejected from AGB stars, provided that the gas is gravitationally trapped within the cluster's potential. We assume that gas particles can be converted into collisionless new stellar particles (`new stars') if the following three physical conditions are met. Firstly, the local dynamical timescale should be shorter than the sound crossing time for a gas particle: this can mimic the Jeans instability of gas. Secondly, the local density ($\rho_{\rm g}$) around a gas particle must exceed a threshold density (${\rho}_{\rm th}$) for star formation:
\begin{equation}
{\rho}_{\rm g} > {\rho}_{\rm th}.
\end{equation}
We assume that this threshold gas density
$\rho_{\rm th}$ is $10^3$ H atoms cm$^{-3}$ for all
models in the present study. Finally, the local velocity field around a gas particle is consistent with that for gravitationally collapsing,
which we model as follows:
\begin{equation}
div {\bf v}<0 .
\end{equation}

\subsection{Gas ejection from AGB stars}

We adopt the same model used in \cite{Bekki_2017Cloud} in order to investigate how AGB ejecta can be accreted onto 1G stellar system. 
In the adopted model, each AGB star ejects gas particles with chemical abundances predicted from recent AGB models (e.g., \cite{Karakas_2010}). The ejection of new particles from AGB stars (`AGB particle') in a simulation means that the total number of gas particles (AGB and ISM particles) can slowly increase as a simulation progresses (after the 1G formation). We consider the ejection of AGB particles at different epochs ($t_{\rm agb}$ depending on the progenitor masses of AGB stars. For example, 
the five epochs of 200, 120, 80, 60, and 40 Myr
correspond to the lifetimes of stars with masses,
(i) $3 \le m_{\rm s} < 4$ (${\rm M}_{\odot}$),
(ii) $4 \le m_{\rm s} < 5$ (${\rm M}_{\odot}$),
(iii) $5 \le m_{\rm s} < 6$ (${\rm M}_{\odot}$),
(iv) $6 \le m_{\rm s} < 7$ (${\rm M}_{\odot}$),
and (v) $7 \le m_{\rm s} <8$ (${\rm M}_{\odot}$),
respectively. We set the minimum mass of AGB stars to $3 {\rm M}_{\odot}$ for three reasons; elemental abundences observed in Galactic GCs require a small age gap between 1G and 2G stars, the fraction of gaseous ejecta from AGB stars with $m_{\rm s}<3 {\rm M}_{\odot}$ is neglidgible and we simulate a short time frame of $<1$ Gyr of evolution.

At the end of a stellar particle's main sequence phase, it ejects an AGB particle with a wind velocity of $v_{\rm wind} = 10$ km s$^{-1}$. The adopted  $v_{\rm wind}= 10$  km s$^{-1}$ is reasonably consistent with recent observations of AGB stars in the LMC (e.g., \citealt{Marshall_etal2004}). The initial temperature of AGB wind ($T_{\rm wind}$) is set to be 1000 K, which is also consistent with standard theoretical models of AGB winds. In low-mass GCs, the initially warm AGB ejecta cannot be converted into new stars owing to the shallow gravitational potentials.

We do not intend on examining the chemical abundances of 1G and 2G stars in the present study for three reasons. Our current simulations do not comprehensively model the ISM abundances within the parent galaxy, and thus in our next paper, we intend on performing an extensive, self-consistent investigation into the initial gas abundances of the GC's host. Secondly, there are some uncertainties in the nucleosynthesis yields of AGB stars in different groups (e.g., \citealt{Karakas_Lattanzio_2014}), which may significantly influence the CNO abundance patterns of 2G stars. Finally, the present study already contains several new results of GC formation, which we wish to focus exclusively on in this paper.

\subsection{Multiple gravitational softening lengths}

In order to investigate gas accretion processes onto GCs within a galactic-scale simulation, we adopt vastly different gravitational softening lengths for dark matter ($\epsilon_{\rm dm}$), stellar and gaseous discs ($\epsilon_{\rm s}$), GCs ($\epsilon_{\rm gc}$). It is challenging to investigate the long-term dynamical evolution driven by
two-body relaxation for massive GCs with $M_{\rm gc} \sim 10^6 {\rm M}_{\odot}$ (1G systems consisting of more than $10^6$ stars) in live galactic gravitational potentials using NBody6 codes, even if there is no gas dynamics (e.g., \citealt{Rossi_etal2016}). Accordingly, we introduce a rather small $\epsilon_{\rm gc}$ for GCs to investigate only the short-term ($\sim 10^8$ yr) gas accretion processes. The long-term evolution of 1G and 2G stars has been demonstrated to be a key process that governs the spatial structures and internal kinematics of GCs with MSPs 
(e.g., \citealt{Vesperini_etal2018}). Clearly, these important long-term dynamical evolution of 1G and 2G stars in GCs should be investigated in future studies with a proper simulation code (e.g., Nbody6).

\subsection{Parameter study}

We mainly describe the results of our fiducial model (M1) 
with $M_{\rm dm}=5 \times 10^{9} \ {\rm M}_{\odot}$,
$R_{\rm vir}=24$ kpc,
$f_{\rm b}=0.036$,
$f_{\rm g}=0.5$,
$M_{\rm s}=9 \times 10^{7} \ {\rm M}_{\odot}$,
$R_{\rm s}=1.7$ kpc,
$M_{\rm gc}=10^{6}\ {\rm M}_{\odot}$,
$R_{\rm gc}=20$ pc,
$f_{\rm rot}=0.3$, inclination angle
$\theta_{\rm gc}=30$ degrees,
and $R_{\rm HI}= 200$ pc ($D_{\rm HI}=400$ pc).
The model parameters for the fiducial model are summarised in Table \ref{tab:model_parameters}. We also investigate models with
different values of these parameters to discuss the observed
diverse physical properties of GCs. The model parameters for these are summarised in Appendix \ref{sec:appendix_tables}.

\begin{table}
\centering
\begin{minipage}{85mm}
\caption{Description of key physical properties  for
the fiducial  model.}
\label{tab:sim_para}
\begin{tabular}{lc}
\hline
{ Parameters } &
{ Values } \\
\hline
Dark matter  mass   & $3 \times 10^{10} {\rm M}_{\odot}$  \\
$c$ parameter  & 16  \\
Virial radius of dark matter halo   & 33.6 kpc  \\
Stellar disc mass   & $9 \times 10^7 \ {\rm M}_{\odot}$   \\
Gaseous disc mass   & $9 \times 10^7 \ {\rm M}_{\odot}$   \\
Stellar disc size   & 2.4 kpc   \\
Gas  disc size   & 2.4 kpc   \\
Mass of 1G stellar system   & $10^6 \ {\rm M}_{\odot}$   \\
Size of 1G stellar system  & 20pc   \\
Mass of a GC-hosting GMC   & $10^7 \ {\rm M}_{\odot}$   \\
HI hole radius  after 1G formation   & 200 pc   \\
Mass resolution (gas)  &  $900 \ {\rm M}_{\odot}$ \\
Size resolution (gas)  &  20 pc \\
Mass resolution (GC)  &  $5 \ {\rm M}_{\odot}$ \\
Size resolution (GC)  &  0.17 pc \\
Number of AGB particles per a gas particle  &  5\\
AGB yield  & \citealt{Karakas_2010} \\
AGB feedback effects  & Included \\
Threshold gas density for star formation  & $10^3$ cm$^{-3}$ \\
\hline
\label{tab:model_parameters}
\end{tabular}
\end{minipage}
\end{table}

\section{Results}
\label{results}
In this work we present an overview of our fiducial model; a 47 Tuc analogue. We analyse the dynamical evolution and accretion history, spatial distribution, kinematics and chemical properties of the 1G and 2G populations. Later, we present our global scaling relations of both cluster and galactic parameters and examine the variety of outputs generated via these simulations.

\subsection{47 Tuc Analogue}
\label{sec:47 Tuc} 

We chose the Galactic GC 47 Tuc as a benchmark for our simulations due to the detailed and comprehensive observational studies available for this cluster. We note that several different combinations of parameters resulted in 47 Tuc-like features (e.g. high 2G mass fraction, evidence of rotation), however, we summarise only one example here. As 47 Tuc is considered to be a metal rich cluster, we adjust the parameters of our host galaxy to be representative of a lower redshift galaxy. We summarise the parameters of the parent galaxies initial conditions in Table \ref{tab:sim_para}. For the GC progenitor, we use a high mass initial 1G capable of generating the expected He and $\alpha$-element enrichment patterns (although numerical quantifications of the elemental composition of the two populations are not considered for our simulation) and a HI hole size relative to the power emitted from SNe from the 1G. After 370 Myr of evolution, the cluster undergoes significant changes, namely, the creation of a 2G. Table \ref{tab:47Tuc} summarises the initial and final values for our fiducial model. We acknowledge that there are significant discrepancies between our stated values and the present-day mass of 47 Tuc. However, the short duration of our simulation leaves $\approx10$ Gyrs of evolution between our masses and that of 47 Tuc. Hence we overestimate both the 1G and 2G masses to account for any dynamical processes which could result in mass loss over this period.

\subsubsection{Dynamical Evolution}
\label{sec:GC_ev}

\begin{figure}
	\includegraphics[width=\columnwidth]{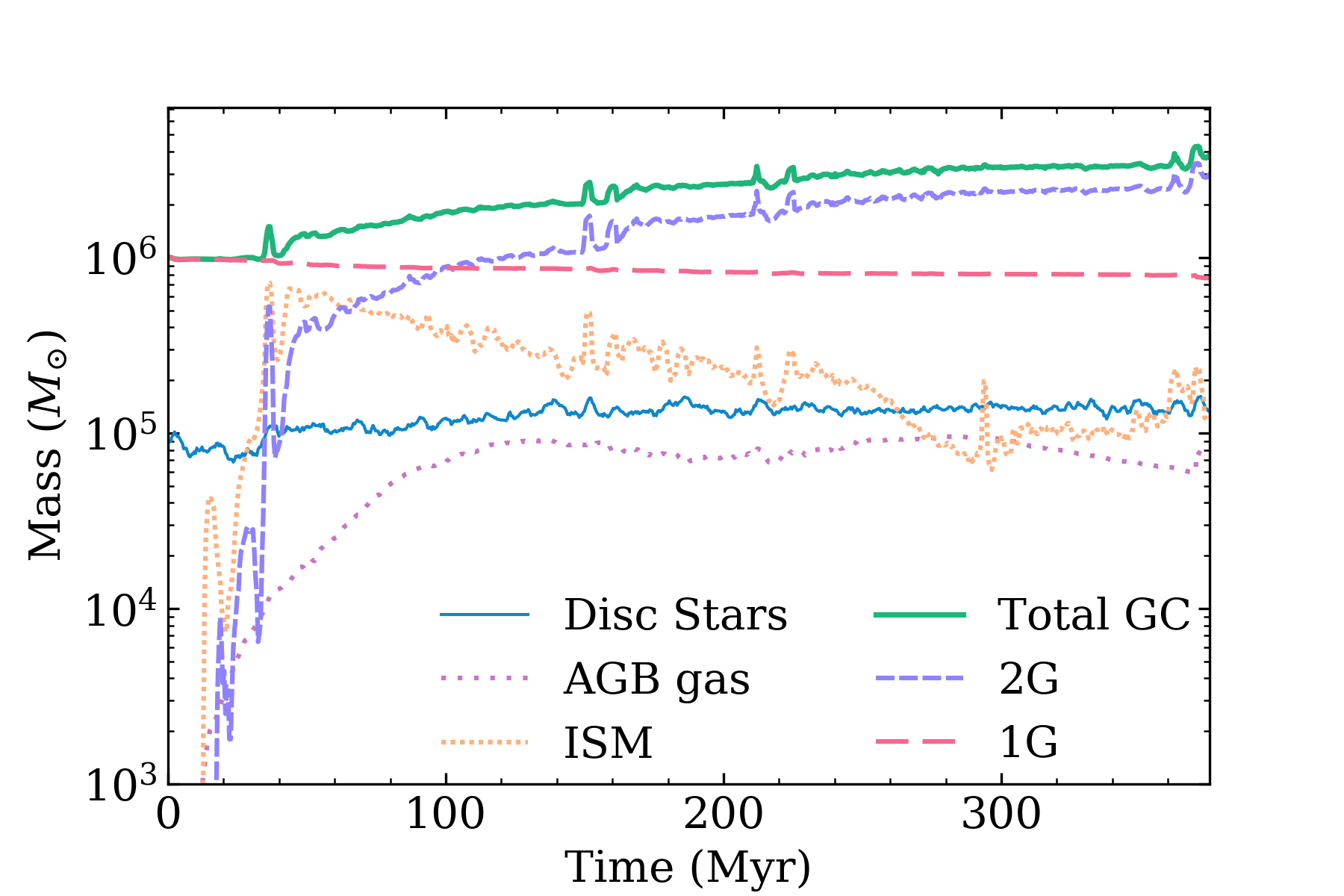}
    \caption{The mass of each component (excluding dark matter) in the simulation as a function of time for the fiducial model. The 1G cluster starts with a set mass of $10^6 \rm{M}_{\odot}$ and is embedded within a $10^5 \rm{M}_{\odot}$ population of disc stars. The HI hole induces a 10 Myr delay in the accretion of ISM from the galaxy. Following the commencement of gas accretion, there is a sharp jump in 2G mass associated with both star formation within the 1G and clumpy accretion. The total GC component represents the sum of the 1G and 2G masses.}
    \label{fig:mass time ev}
\end{figure}

Here we describe the dynamical evolution of a GC (47 Tuc analogue) after its formation in an exponential gas disc with very clumpy ISM. The simulation runs in two phases. In the first phase, we evolve a galaxy with an exponential disc for a period of 100 Myr. This is to create gas density perturbations to select possible cluster progenitors. After this time period, a high density gas clump and its surrounding region is replaced with a cluster to approximate the formation of the 1G. Apart from the selection of a high density region, we do not perform any further analysis on this phase in the present investigation. The second phase is the focus of our study and henceforth we use T = 0 to represent the start of this simulation. We place the 1G in the previous location of the high density GMC and create a $R_{\rm{HI}}$=200 pc HI hole to emulate SNe effects of the 1G. Thereafter, the simulation records the evolution of various parameters within a radius of 30 pc from the centre of the 1G (1.5$\times R_{\rm{gc}}$). Fig. \ref{fig:mass time ev} plots the evolution of 1G, 2G and the sum of these two components to give the total mass of the GC. We decompose the gas particles into two separate components; ISM and AGB gas. ISM is representative of pristine gas originating from the parent galaxy whereas AGB ejecta is enriched gas released from stars during the AGB phase. Disc stars originate from the galactic disc but have settled within the cluster's potential.

At time T = 0, there is a $10^6 \rm{M}_{\odot}$ GC progenitor (as stated in Table. \ref{tab:47Tuc}) and $10^5 \rm{M}_{\odot}$ of disc stars within the radius of the cluster. The 200 pc HI hole causes a $\approx10$ Myr delay before the GC starts to interact with gas from the galaxy. Gravitational attraction is the primary mechanism by which the 1G accumulates the ISM. At a similar time, AGB stars release low-density gas into the GC. As the 1G stars are the source of the AGB ejecta, this gas accumulates in the centre of the cluster whereas the ISM particles distributes itself at a larger radii. Within 50 Myr, the mass of the 2G is over 50\% of its final value. The initial spikes in 2G mass (e.g. at T=35 Myr, T=45 Myr ) are due to what we have termed `clumpy accretion'. Clumpy accretion occurs when new stars form externally to the GC but fall into its gravitational potential and mix with the existing population. We observe this phenomenon in most simulations, so we discuss the impact on the chemistry of the two populations later in this work. In the first 50 Myr, the mass of the 2G reflects the mass of the ISM, which could illustrate a coupling between the ISM and the formation of new stars. Later, the introduction of AGB gas becomes significant and the mixture between pristine and enriched gas creates the remainder of the 2G stars. Disc stars from the galaxy play a minor role in the dynamical evolution of the GC; maintaining a relatively constant mass of  $10^5 \rm{M}_{\odot}$. However, if the gravitational potential of the cluster trapped these disc stars, it may increase the mass of the GC after 350 Myr by $\sim 3\%$ based on the GCs 20 pc radius. These disc stars could be a potential candidate for the older population of 47 Tuc seen though the photometry of the sub-giant branch (see \citealt{milone_etal2012}). The mass of disc stars within the cluster after the long term dynamical evolution is still an open question. We intend to investigate this as well as the probability of natal GMC's capturing disc stars in future works.

\begin{figure}
	\includegraphics[width=\columnwidth]{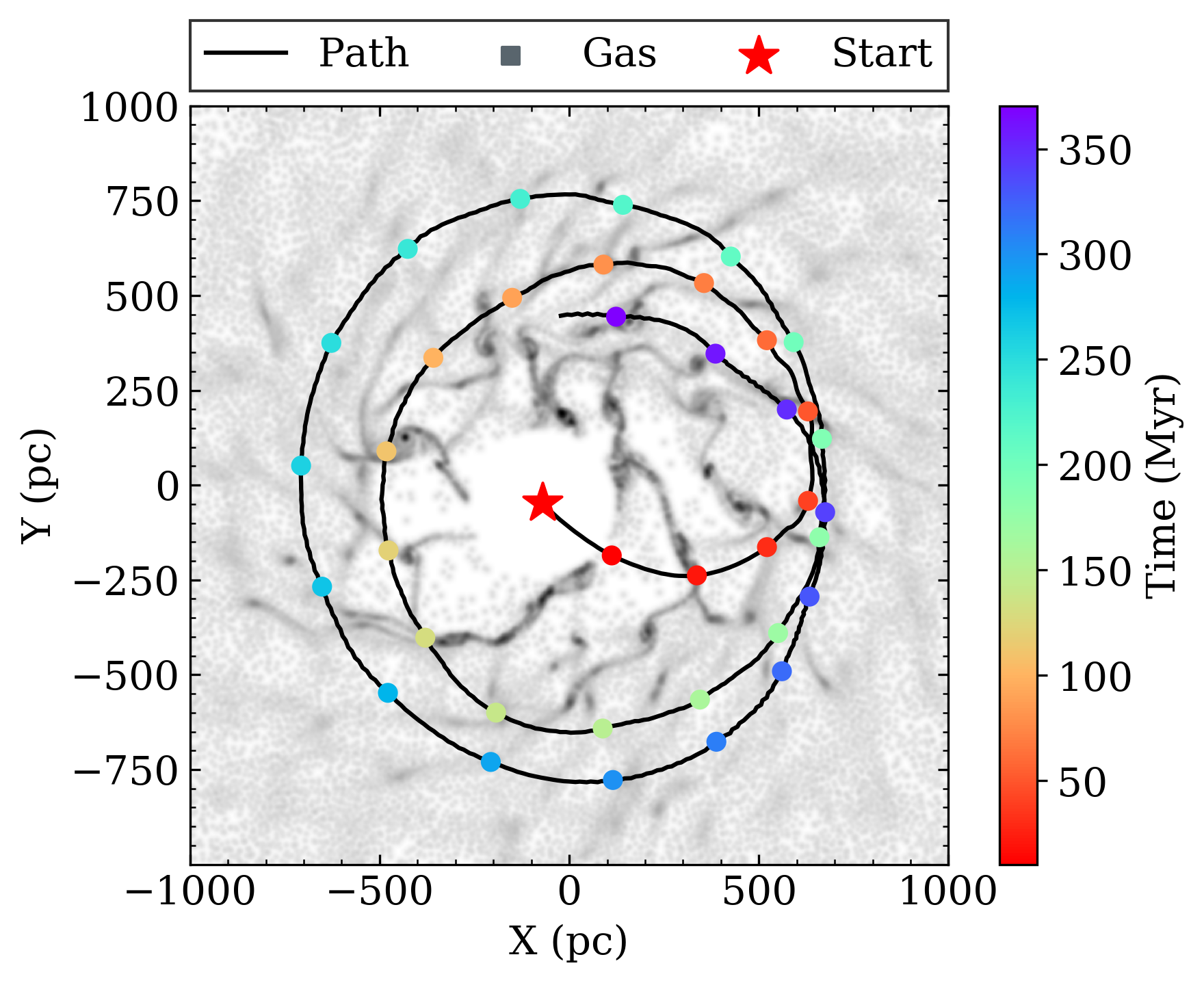}
    \caption{The orbital evolution of the cluster from T = 0 to T = 370 Myr. The coloured dots are placed at 10 Myr intervals and range from red at T = 0 to blue after 370 Myr. The gas surface density distribution of the host galaxy at the start of the simulation has been drawn under the orbital trajectory for reference.}
    \label{fig:orbit time ev}
\end{figure}

In Fig. \ref{fig:orbit time ev}, we plot the orbital path of the GC progenitor within its host galaxy. The X-Y axis defines the spatial extent of the clusters traversal of the galaxy, with the path taken by the GC shown with a solid black line. A red star delineates the clusters starting position, and we place a rainbow point at 10 Myr intervals in order to identify the clusters location at a specific time. The cluster starts $\approx85$ pc from the galactic centre and has an initial velocity of $26 \ \rm{kms}^{-1}$. The main accretion events occur from 20 to 50 Myr (i.e. Fig. \ref{fig:mass time ev}). During this time, the progenitor cluster experiences a substantial increase in mass and interacts with neighbouring high density regions. A combination of centrifugal forces and sling-shot effects from neighbouring clusters forces the GC progenitor into a larger orbit. The cluster settles into a stable orbit and reaches an apocenter of $\approx700$ pc, after 300 Myrs. We plot the initial distribution of gas to illustrate the various density perturbations as well as the scale of the HI hole. The high density filaments of gas are the primary source of fuel for future star formation within the cluster.

\begin{figure*}
	\includegraphics[width=\textwidth]{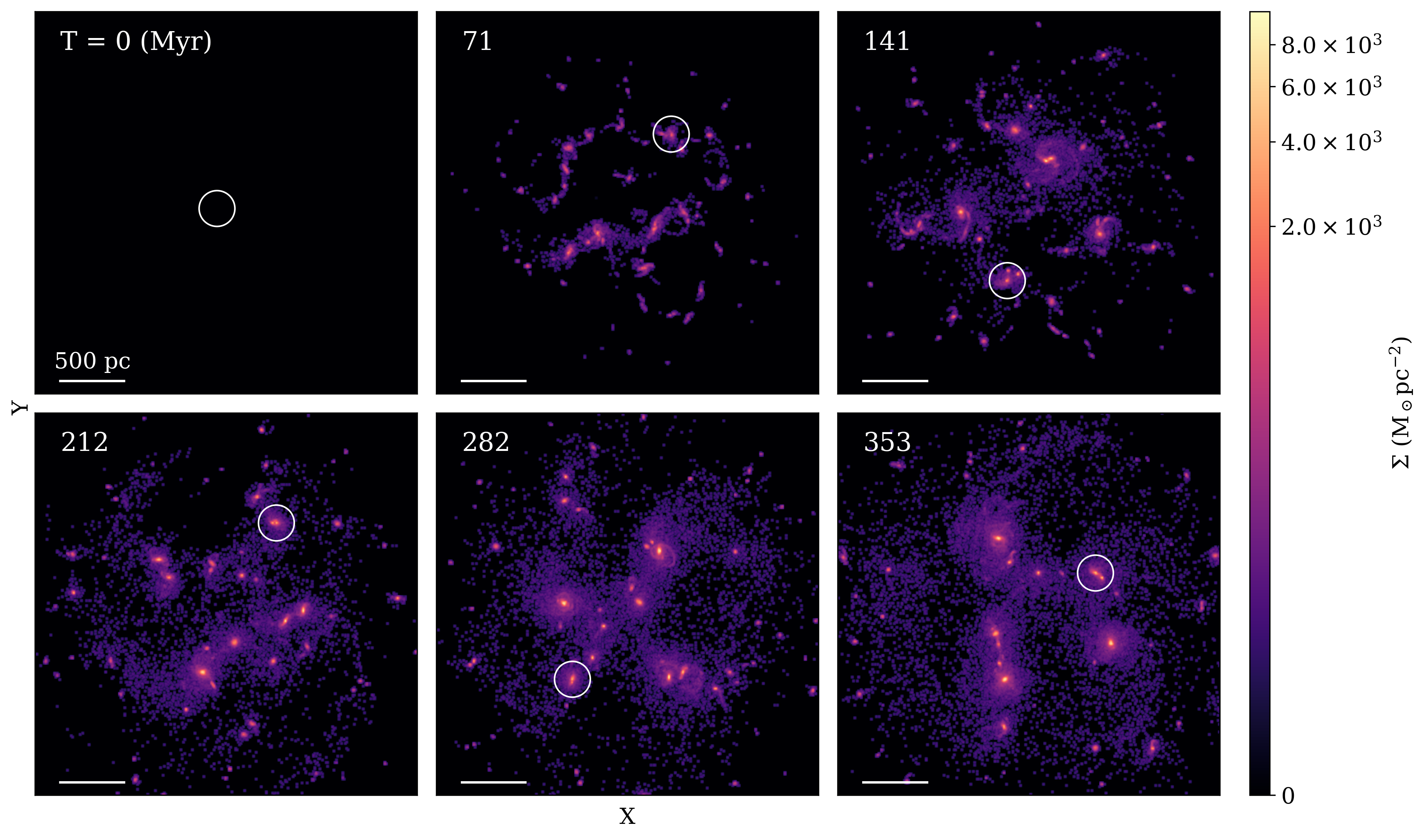}
    \caption{The spatial distribution of new stars which formed during the simulation. The X-Y projection spans a 3 kpc range, centred on the parent galaxy. The text in the top left hand corner denotes the time at which that snapshot was taken. In each of the panels, a white circle denotes the location of the 1G progenitor at each time step. After 71 Myr, a number of low mass clusters emerge out of high density gas perturbations and these continue to evolve throughout the duration of the simulation.
    The corresponding 1G, gas and disc star figures are located in Appendix \ref{sec:spatial_comp_app}.
    }
    \label{fig:time_ev_new}
\end{figure*}

Complementing the previous plots, Fig. \ref{fig:time_ev_new} presents the evolution of new stars which form during the simulation. The first time step in the top left corner contains a white circle which denotes the starting location of the 1G. The next time step occurs after the collapse of the HI hole where the 1G has established a population of stars from a mixture of AGB ejecta and pristine gas. We see this population at 71 Myr in the top centre panel, with the white circle encompassing a small population of new stars which make up the foundations of the 2G. The remainder of the panels show this population evolving through the simulation.

The initial distribution of both the GC and the host galaxy greatly influence the development of the GC. The companion plots to Fig. \ref{fig:time_ev_new} in Appendix \ref{sec:spatial_comp_app} illustrate the dynamics of gas particles and disc stars. The first phase of the simulation generates the turbulent patterns which spawn GMCs observed in Fig. \ref{fig:time_ev_gas} and allows us to select a gas cloud with mass $> 10^7 \rm{M}_{\odot}$. At T = 0, high density regions of gas and 2G stars coincides with over-densities of disc stars. Fig. \ref{fig:mass time ev} indicates that the total mass of disc stars trapped at the commencement of the simulation is $\approx 10^5 \rm{M}_{\odot}$. Both Fig. \ref{fig:mass time ev} and Fig. \ref{fig:time_ev_gal} demonstrate that this mass remains relatively constant over the next 370 Myr. Our simulation provides evidence that GMCs which form within a galaxy may have the potential to capture a population of disc stars before the 1G of a GC is established. Furthermore, these older disc stars may constitute the metal poor anomalous population observed in some Galactic GCs (e.g. Terzan 5; \citealt{Ferraro_etal2009}). We discuss this further in Section \ref{sec:disc stars} and intend for it to be the focus of forthcoming papers in this series.

In the present simulations, the first phase of evolution restricts gas particles from transforming into new stars. However, more physical simulations would allow for high density regions of gas to transform into disc stars. With this prescription implemented however, the outcome of these simulations were nearly identical to that of our fiducial model.

\begin{table}
	\centering
	\caption{The parameters at the start and end of the 370 Myr evolution of our fiducial model. This model was chosen as a potential candidate for modelling the evolution of the Galactic GC 47 Tuc. The discrepancy between these masses and 47 Tuc's present day mass is to account for the $\approx10$ Gyrs of evolution after the creation of the 2G.}
	\label{tab:47Tuc}
	\begin{tabular}{lcccc}
	    \hline
        &$\mspace{90mu}$ & Initial  &$\mspace{30mu}$ & Final \\ 
        \hline
        $\rm{M}_{1G}< R_{5}$  $(\rm{M}_{\odot})$ & & 5.07$\times 10^5$ & & 3.95$\times 10^5$ \\ 
        $\rm{M}_{1G}< R_{10}$ &  &8.51$\times 10^5$ & & 6.47$\times 10^5$ \\ 
        $\rm{M}_{1G}< R_{20}$ &  &$10^6$ & & 7.37$\times 10^5$ \\ 
        $\rm{M}_{2G}< R_{5}$ &  &- & & 4.31$\times 10^5$ \\ 
        $\rm{M}_{2G}< R_{10}$ &  &- & & 1.10$\times 10^6$ \\ 
        $\rm{M}_{2G}< R_{20}$ &  &- & & 2.19$\times 10^6$ \\ 
        $\rm{V}_{x, \ 1G}$  $(\rm{km\ s}^{-1})$ &  &-3.77 & & -1.65 \\ 
        $\rm{V}_{y, \ 1G}$ &  &3.77 & & -1.66 \\ 
        $\rm{V}_{x, \ 1G}$ &  &- & & 0.03 \\ 
        $\rm{V}_{x, \ 2G}$ &  &- & & -9.02 \\ 
        $\rm{V}_{y, \ 2G}$ &  &- & & 6.98 \\ 
        $\rm{V}_{x, \ 2G}$ &  &- & & -0.31 \\ 
        $\sigma_{x, \ 1G}$ &  &9.62 & & 11.04 \\ 
        $\sigma_{y, \ 1G}$ &  &9.63 & & 11.07 \\ 
        $\sigma_{z, \ 1G}$ &  &10.51 & & 10.84 \\ 
        $\sigma_{x, \ 2G}$ &  &- & & 6.47 \\ 
        $\sigma_{y, \ 2G}$ &  &- & & 5.34 \\ 
        $\sigma_{z, \ 2G}$ &  &- & & 2.61 \\ 
 
		\hline
	\end{tabular}
\end{table}

\subsubsection{Morphological Properties}
\label{sec:morphology}

\begin{figure*}
	\includegraphics[width=\textwidth]{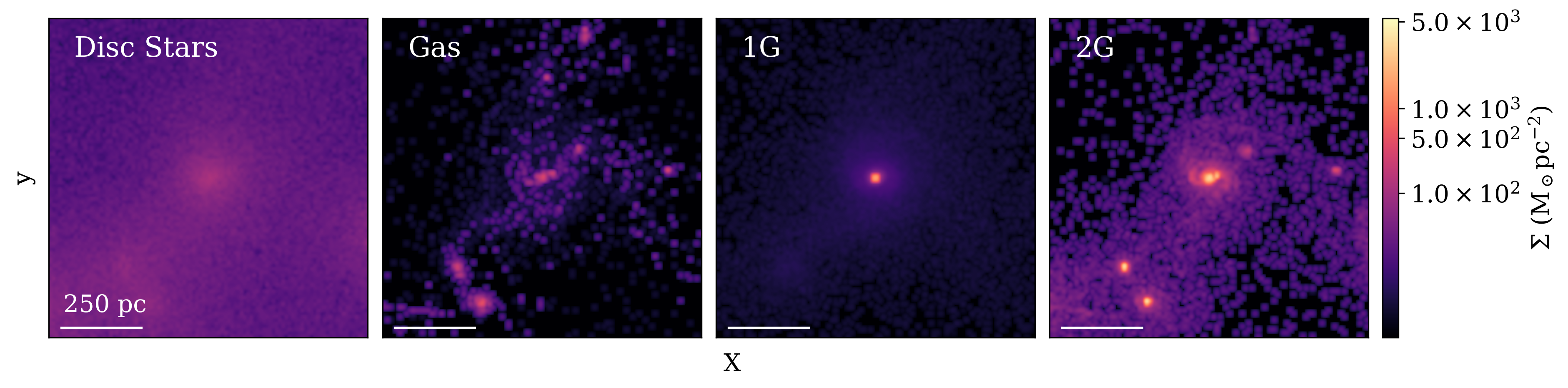}
    \caption{The spatial density relation for the 1G, 2G, gas and disc stars after 370 Myr of the simulation. The 1 kpc box shows the complex structure surrounding the GC. High-density clusters are visible in the bottom left corner for the 2G, gas and disc star components.}
    \label{fig:final_ts}
\end{figure*}

Fig. \ref{fig:final_ts} shows the final surface density distributions of the 1G, 2G, gas and disc stars. As expected, the 1G component appears as a single, high-density point within the simulation box. The 2G component has fragmented into multiple clumps with a variety of sizes and masses. The density perturbations in the gas component, which represents both the ISM and AGB gas, mirrors the distribution of the 1G and 2G particles. The panel depicting the gas component also illustrates that there is a non-negligible mass of gas remaining within the cluster. As present-day GCs are known to be incredibly gas-poor, there must be some mechanism for expelling this gas during the remainder of the cluster's evolution. Possible mechanisms are briefly discussed in Section \ref{sec:2G_concentration}. The over density of disc stars seen in the very first time step still surround the cluster. The 2G population has fragmented into multiple clusters with comparable sizes and densities to our fiducial cluster. We observe clusters in the lower left quadrant of the gas and 2G panels undergoing a merging event. Additionally, several smaller clusters are in the process of being accreted into larger ones. This points to a hierarchical nature of the accretion process. 

\begin{figure}
	\includegraphics[width=\columnwidth]{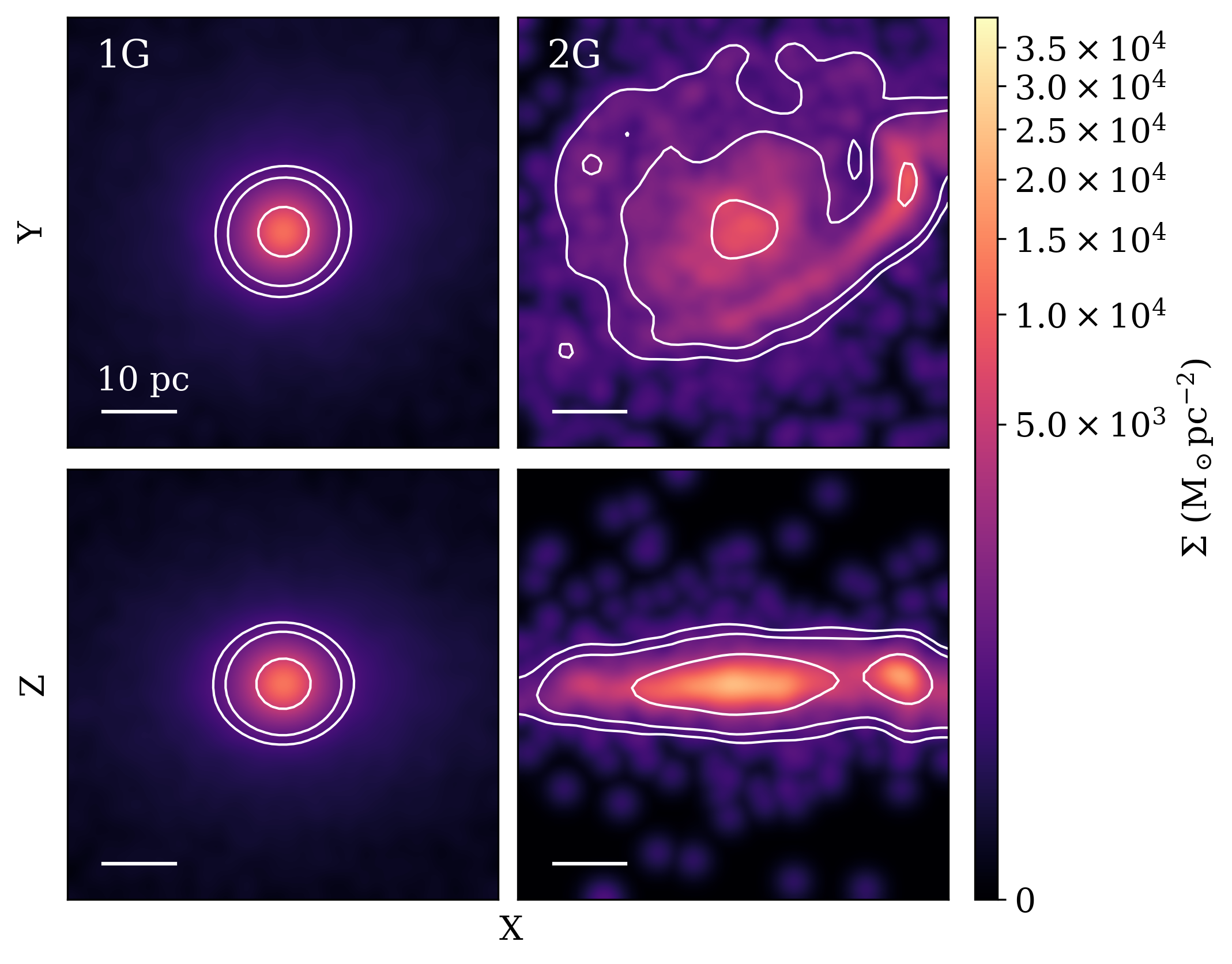}
    \caption{The X-Y and X-Z distributions of the 1G and 2G populations. Smoothed contour lines are drawn at $5\times 10^2 \rm{M}_{\odot}$, $1\times 10^3 \rm{M}_{\odot}$ and $5\times 10^3\rm{M}_{\odot}$ to emphasise the asymmetry of the 2G. Both the distribution and contours have been smoothed using a Gaussian filter function.}
    \label{fig:time_ev_final_GC}
\end{figure}

The structure and distribution of the 1G and 2G are well documented in the literature. A well known criteria of theoretical models of multiple population formation is that the 2G is more centrally concentrated than the 1G. We investigate this for our models by looking at the morphologies of the 1G and 2G in a 40 pc box surrounding the GC. Fig. \ref{fig:time_ev_final_GC} is centred on the 1G's centre of mass and illustrates the spatial extent of the 1G and 2G components. The majority of the 1G's mass is localised within a 10 pc radius with contours drawn at $5\times 10^2 \rm{M}_{\odot}$, $1\times 10^3 \rm{M}_{\odot}$ and $5\times 10^3\rm{M}_{\odot}$. From the X-Z projection, the 1G has been slightly elongated due to its angular momentum but is approximately spherically symmetric. In the X-Y projection of the 2G stars, the central isophote at $5\times 10^3\rm{M}_{\odot}$ occurs at an equivalent radii to the 1G. However the $5\times 10^2 \rm{M}_{\odot}$ and $1\times 10^3 \rm{M}_{\odot}$ isophotes illustrate the complex distribution of 2G stars. Firstly, the extended structure located on the right side of the figure is indicative of interactions with ex-situ 2G stars from the environment. The gravity from the 1G and 2G structures has the potential to tidally attract and disrupt smaller 2G clusters within the parent galaxy. The spatial extent of the 2G is much larger than the 1G. One striking difference between the 1G and 2G is the disc like structure of the 2G seen in the X-Z projections. The disc forms parallel to the plane of the galaxy as a result of its angular momentum. \cite{Mastrobuono-Battisti_Perets_2016} investigated the long-term evolutionary effects of 2G stellar discs in dense star clusters using N-body simulations and found that as the 2G disc relaxes, the 1G stellar population flattens the GC structure to become more elliptical. In our simulations however, the 2G disc extends far beyond the radius of the 1G and could potentially be tidally stripped before the relaxation process is complete. We intend to study the long term evolution of this system in future works to investigate the final distribution of the 1G and 2G stars within the cluster. The colour map shows that the central regions of the 1G and 2G have comparable densities. Equal central densities are observed in some GCs, however \cite{milone_etal2012} showed that the central 5 pc had a ratio of $\approx0.8$ 2G stars. This ratio decreased to $\approx0.6$ within 15 pc (conversions from arc minutes to parsecs used the distance values from \citealt{Harris_1996}(2010 edition) ). Given the unrelaxed nature of the 2G, future studies will test whether this extended halo of 2G stars collapses into the centre, thus causing a higher concentration akin to 47 Tuc.

\begin{figure}
	\includegraphics[width=\columnwidth]{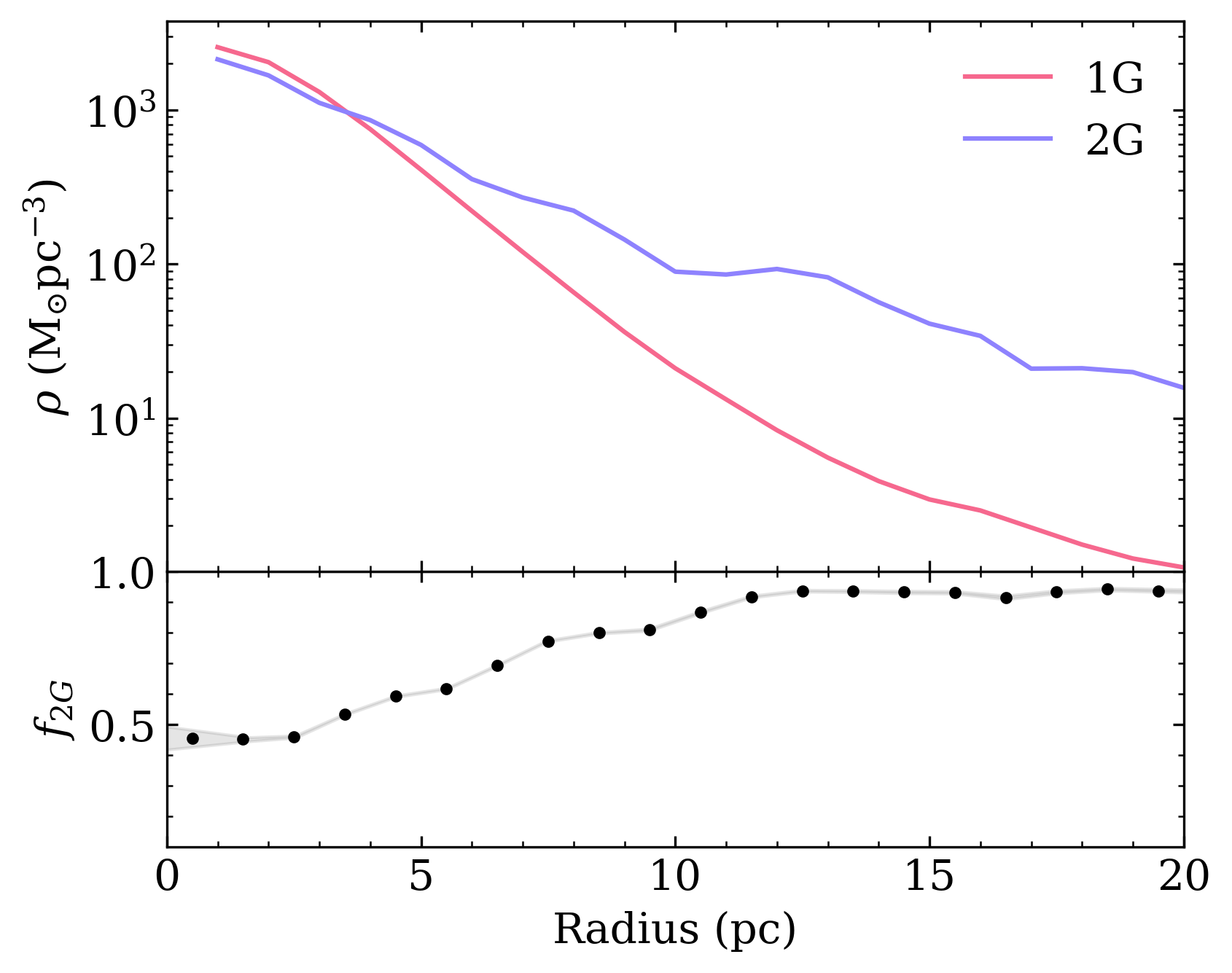}
    \caption{The 3D radial density distributions for the 1G and 2G stars and corresponding fraction of 2G stars. In the central 5 pc of the cluster, the 1G and 2G densities are approximately equivalent. However at higher radii, the 2G becomes the dominant source of mass within the cluster. The shaded region represents the $1\sigma$ standard deviation.}
    \label{fig:f_enriched}
\end{figure}

As the newly formed GC system is in a highly disturbed state, it would be naive to compare observational radial distributions directly to our results. However, we include Fig. \ref{fig:f_enriched} for completeness. As demonstrated by Fig. \ref{fig:time_ev_final_GC}, the central concentrations between the 1G and 2G are comparable in mass. Fig. \ref{fig:f_enriched} plots the fraction of 2G stars, $f_{2G}$ out to a radius of 20 pc. Within the central 5 pc, the fraction of 2G stars (i.e. $\frac{M_{2G}}{M_{2G}+M_{1G}}$) is slightly below $\approx$ 50\%. As the density of the 1G decreases, the extended nature of the 2G results in the outer regions of the GC being entirely dominated by the 2G. In their current state, these results align with observations from some clusters which have a centrally concentrated 1G but not with those from 47 Tuc. Again we highlight that our analysis is based on a newly born 2G and that its evolution across time may significantly affect its spatial distribution within the cluster. The results of evolving this system further in time will be presented in future studies.

One repercussion of Fig. \ref{fig:f_enriched} is the prediction that very young GCs should have a colour gradient. Newly formed GCs should become bluer in colour at increasing radii due to the higher fraction of younger stars. This appears counter-intuitive as 2G formation should be predominantly taking place within the centre of the 1G. We intend to investigate the timescale in which these stars are either accreted or tidally destroyed by the host galaxy in future studies of the long term evolution of the cluster. Observational confirmation of any colour gradients in high redshift clusters may be possible in the future using high resolution photometric and spectroscopic instruments.

\subsubsection{Kinematic Properties}
\label{sec:kinematics}

\begin{figure}
	\includegraphics[width=\columnwidth]{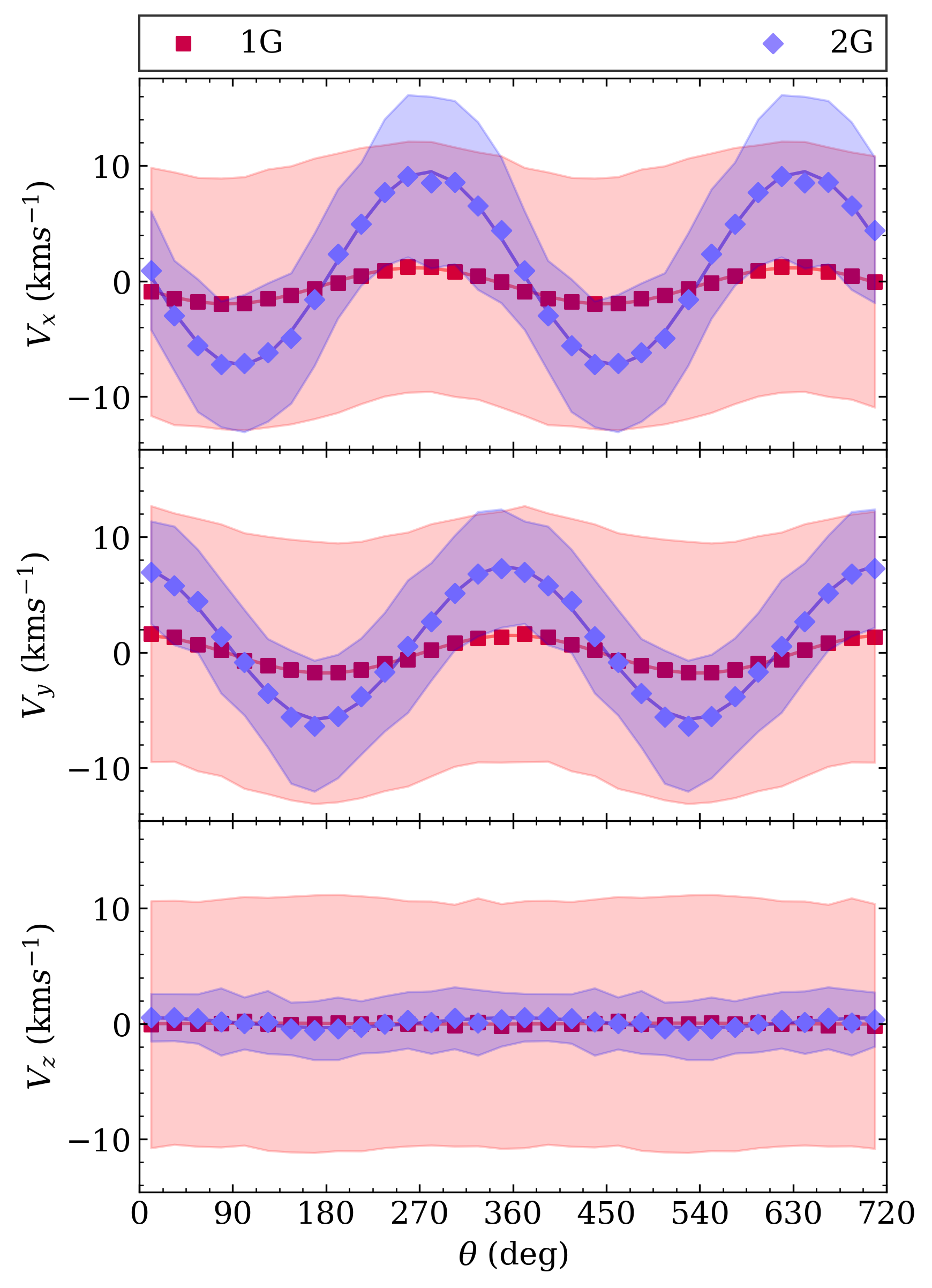}
    \caption{Rotation amplitude as a function of $\theta$ for the fiducial model. The red and blue shaded regions corresponds to the $1\sigma$ velocity dispersion. The amplitude of the 2G is far more significant than the amplitude of the 1G. The GC is rotating around the z-axis and thus we do not expect any rotation within this component.}
    \label{fig:polar_rot}
\end{figure}

Primarily driven by Gaia DR2, kinematical studies of GCs have recently experienced a growth in popularity. As such, we investigate the kinematics and internal rotation of our fiducial model. The simulation includes the option to induce rotation into the 1G population. Additionally, the GC can build up angular momentum from the accretion of both stellar clumps and gas, as well as gaining some residual rotation from its orbit around its galaxy. As evident from Table \ref{tab:47Tuc}, the process of accreting new material onto the progenitor cluster drastically alters its kinematics. \cite{Milone_etal201847Tuc} used Gaia DR2 to analyse the rotation of the 1G and 2G for 47 Tuc and found that the rotation of the two populations was in-phase and had similar amplitudes. Fig. \ref{fig:polar_rot} plots the amplitude of velocity in the x, y and z directions as a function of the angle from positive X axis, $\theta$, for our simulated 47 Tuc analogue. The velocities were obtained by dividing each population into 16 equal-sized bins ranging from $\theta$ = 0 to 360. The average x, y and z velocity were calculated for each bin and plotted against the corresponding mean value of $\theta$. The shaded region illustrates the velocity dispersion for each bin. Next, we apply a sinusoidal curve to each population and compare the derived amplitudes. The rotation of our two populations is in phase, which is in agreement with observations of 47 Tuc. However, the amplitude of the 2G is much larger than that of the 1G. Furthermore, as demonstrated by the shaded regions denoting the velocity dispersion, the 2G exhibits far more coherent rotation compared to the 1G, which is at odds with most observations of GCs. In a study of M80, \cite{Kamann_etal2020} found that the more nitrogen-enriched population rotates faster than the other two populations which aligns with our results. Additionally, the aforementioned study by \cite{Mastrobuono-Battisti_Perets_2016} states that their 2G stellar population is characterised by a lower velocity dispersion and a higher rotational velocity compared with the primordial older population which is echoed by these results. In our future works investigating the long term evolution of the cluster, we intend to explore whether the differences between the rotational velocity of the 1G and 2G diminishes as to align with current observations of Galactic GCs. Using a suite of N-body simulations, \cite{Henault-Brunet_etal2015} determined that different pollution scenarios could result in distinct kinematic properties which could be used to distinguish between various scenarios. Similarly, we intend to analyse the kinematic imprint of the different rotation amplitudes between each population for our scenario, the outcome of which could either strengthen or falsify our model.

\begin{figure}
	\includegraphics[width=\columnwidth]{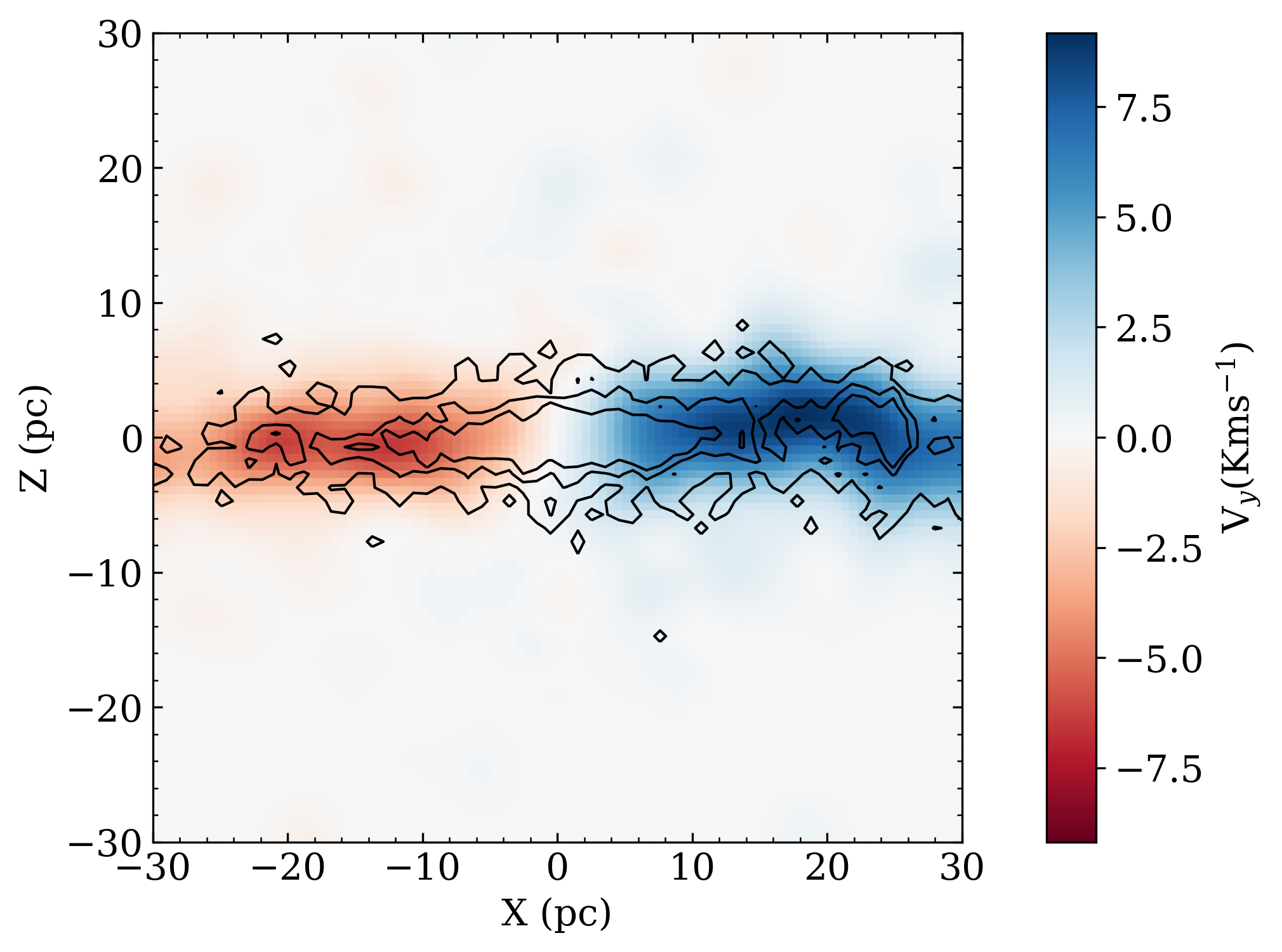}
    \caption{The velocity map along the line of sight for the 2G. Contours are placed at densities of $1\times 10^3 \rm{M}_{\odot}$, $5\times 10^3 \rm{M}_{\odot}$ and $1\times 10^4\rm{M}_{\odot}$ to show the spatial extent of the stars. The rotation map has undergone Gaussian smoothing however the contours are as-is to show additional structure within the plane of the 2G disc.}
    \label{fig:v_map}
\end{figure}

Turning our attention to the rotation of the 2G only, Fig. \ref{fig:v_map} shows the velocity map along the line of sight. A Gaussian filter has been applied to smooth over any outlying escaping 2G star particles. The contours serve as a reference for the mass isophotes. The 2G disc is clearly rotating along the line of site with a maximum velocity in the $y$ direction of $\approx 8 \rm{kms}^{-1}$. The high velocity dispersion and low rotation of the 1G cause any signs of rotation to be far less obvious and thus this plot is omitted from the present study.

\subsubsection{Chemical Enrichment}
\label{sec:chem_enrich}

Studying chemical compositions in GCs is central to their characterisation. Clusters which are homogeneous in heavy element abundances are known as Type I GCs whereas Type II GCs are those with more than two dominant populations. Our simulations can reproduce both types of GCs through altering the time, mass and starting location of any accreted clumps.

\begin{figure}
    \includegraphics[width=\columnwidth]{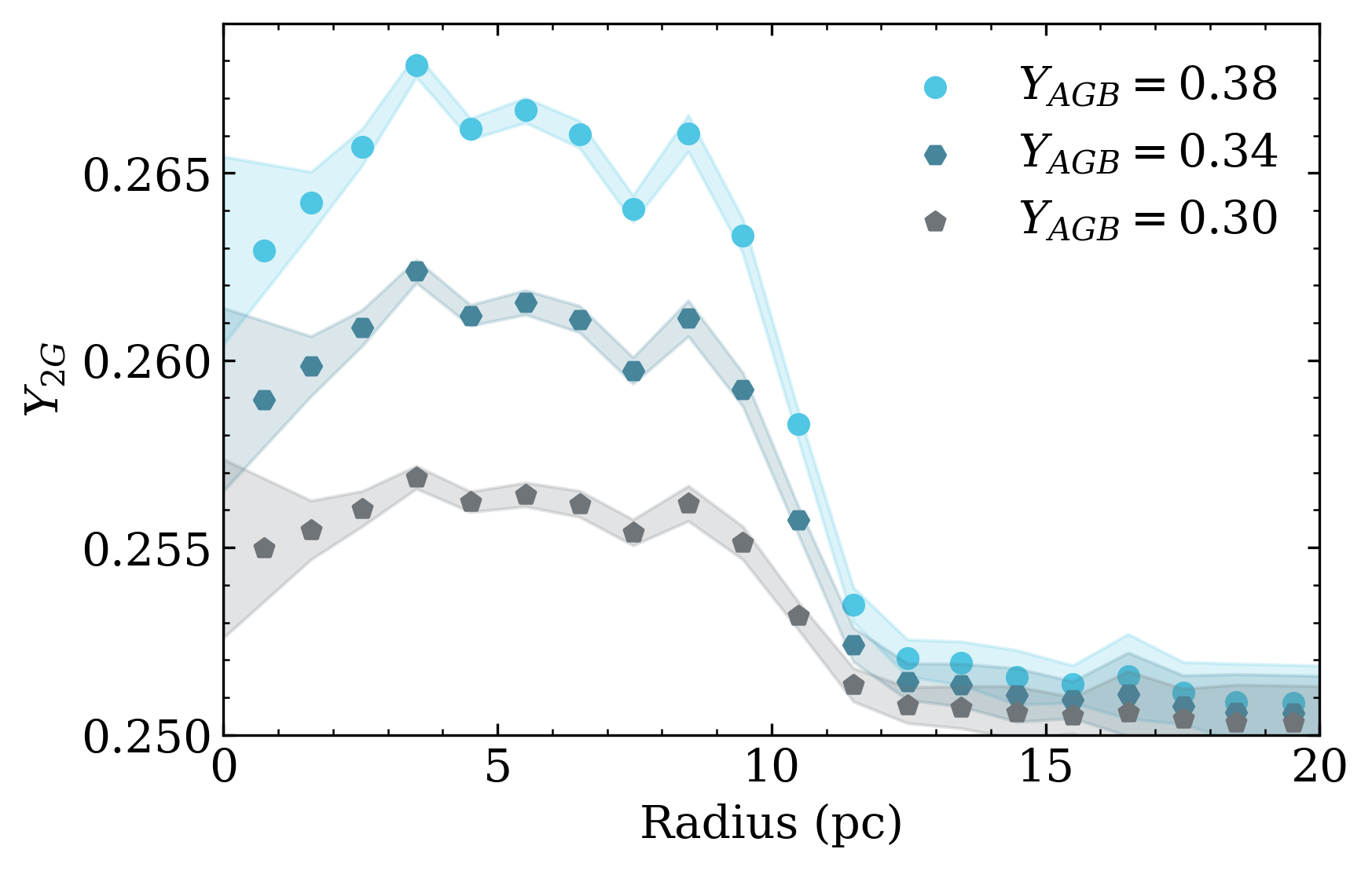}
    \caption{The helium abundance for the 2G for three different $Y_{AGB}$ values. The shaded regions represent the $1\sigma$ standard deviation. The transition to the helium abundance of pristine gas ($Y = 0.25$) corresponds to the radial extent of the 1G.}
    \label{fig:he}
\end{figure}

One of the main strengths of the AGB scenario is its ability to explain the Mg-Al anti-correlation and other elemental patterns, as well as being able to deposit a significant mass of Helium (He) into the GC. We estimate the He enrichment of the 2G via the formula:
\begin{equation}
    Y_{2G} = \frac{ Y_{AGB}\times M_{AGB} + Y_{ISM} \times M_{ISM}}{M_{AGB}+M_{ISM}}
\end{equation}
where  $Y_{AGB}$ is the typical He abundance for AGB ejecta and $Y_{ISM}$ is the He abundance for pristine gas in the ISM. $M_{AGB}$ and $M_{ISM}$ are the masses of new stars formed from AGB ejecta and pristine ISM respectively. Fig. \ref{fig:he} shows the He enrichment as a function of radius with the shaded regions corresponding to the $1\sigma$ standard deviation. We assume three different He abundances in AGB gas; $Y_{AGB} = $ 0.3, 0.34 and 0.38, and pristine ISM to be $Y_{ISM} = 0.25$. Stars in the central 9 pc are He enriched, with $Y_{2G} > Y_{ISM}$ for all He abundances. This is expected as this region has a high density of 1G stars (Fig \ref{fig:f_enriched}) which are the primary source of AGB ejecta. After 9 pc, there is a steep decline in the He enrichment and at radii greater than 12 pc, the He abundance drops to approximately that of $Y_{ISM}$. This calculation assumes that $Y_{ISM}$ is constant throughout the galaxy when more realistically, the GC could gravitationally interact with inwardly or outwardly moving gas in the disk. This could allow the GC to accrete gas with various $Y_{ISM}$, generating a larger $\Delta Y$ between the 1G and 2G.

An early study of 47 Tuc by \cite{Briley_1997} suggested that a non-negligible spread in the He abundance may explain the morphology of the horizontal branch. Later, \cite{Anderson_etal2009} required a $\Delta Y = 0.026$ in order to explain the colour spread in the main sequence, provided that He is the main source of broadening. Today, the observed He abundance of 47 Tuc is given to be $Y_{1G}=0.256$ and $Y_{2G}=0.276$ \citep{Fu_etal2018}. Super AGB stars (e.g., \citealt{Siess_2010}; \citealt{Ventura_DAntona2011}; \citealt{Doherty_etal2014}) may be capable of producing a He abundance of $0.38$, however this is still not enough to generate the $\Delta Y$ observed for 47 Tuc. Approaches to this problem include increasing the He concentration of the pristine gas to match the 1G or decreasing the dilution of AGB gas by limiting ISM accretion. Additionally, due to the uncertainties around AGB yields, future revisions of this model with updated AGB prescriptions may be able to rectify this difference.

\begin{figure}
    \includegraphics[width=\columnwidth]{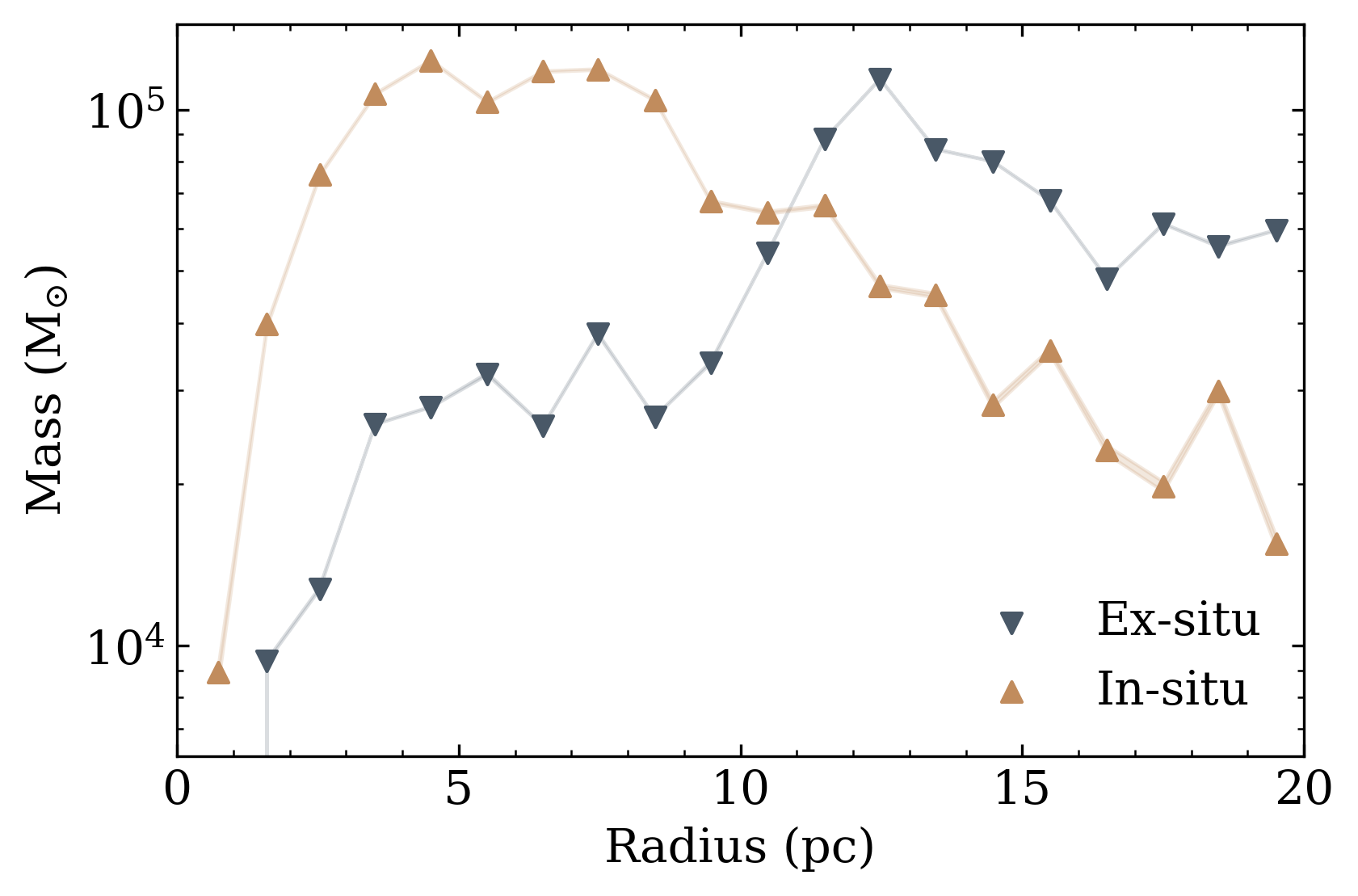}
    \caption{The radial distribution of in-situ and ex-situ stars within 1 pc bins. Stars formed in-situ to the 1G make up the majority of the central region of the 2G. Stars which formed externally to the 1G make up the dominant population outside the 10 pc radius.}
    \label{fig:in_ex_situ}
\end{figure}

The occurrence of in-situ and ex-situ stars has yet do be discussed in the context of GC evolution. Clumpy accretion within proto-GCs naturally explains Type II GCs, but the chemical consistency of Type I GCs may pose some issues for our scenario. Furthermore, there is no guarantee that these new stars are genuine members of the 2G population and may even have chemistry representative of the 1G. Given the degree of He enhancement of the 2G within the central region of the GC, we would expect that a significant proportion of these stars were formed within the GC environment. We confirm this hypothesis in Fig. \ref{fig:in_ex_situ}, which plots the mass of in-situ and ex-situ stars at increasing radii. Within the central ten pc which we know is dominated by the 1G, the majority of the stars formed in-situ with the GC. With increasing radius, the dominant population of 2G stars becomes those which were formed outside the 1G, and were assimilated into the GC at a later time during a clumpy accretion event.

\begin{figure}
    \includegraphics[width=\columnwidth]{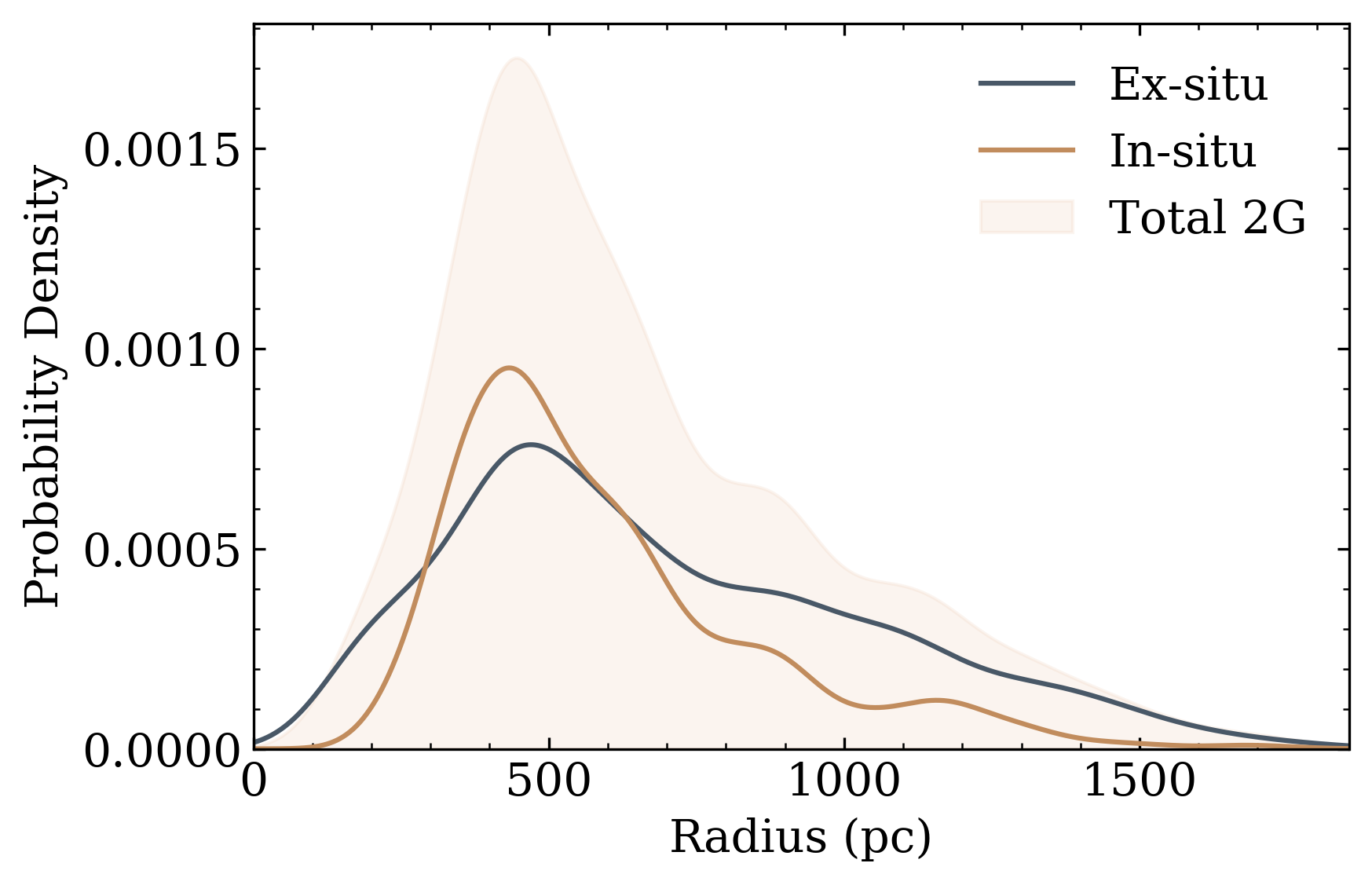}
    \caption{The probability density plot of the original location of new stars. The in-situ and ex-situ fractions have been scaled according to their total contribution to the total 2G. The original locations of each type of new star peaks at the same location and thus should have similar metallicities.}
    \label{fig:in_ex}
\end{figure}

\par
\indent
In addition to variations in He, Lithium (Li) abundances also poses a challenge to many MSP formation scenarios (for a recent discussion, see \citealt{Gratton_etal2019}). Forged during the Big Bang nucleosynthesis, Li may be vital in determining the stellar source of GC polluters. The AGB scenario has been heralded as a promising solution to the problem of Li variations between populations due to AGB stars ability to manufacture Li through the Cameron Fowler mechanism \citep{Cameron_Fowler_1971}. However, the abundance of Li in pristine material accreted onto the GC has been suggested to influence this result \citep{DAntona_etal2012}. Here we do not consider the influence of lithium production and dilution effects due to the vast uncertainties from AGB yields; however, our model's ability to generate a diverse range of 2G stars enriched by various quantities of AGB gas works in our favour.

The metallicity gradient of galaxies is well documented in the literature (e.g., \citealt{Searle_1971}; \citealt{Magrini_etal2016}; \citealt{Bresolin_2019}). We assume that if the accreted clump originated from a similar location within the GC, it would have similar metalicities to the accreted ISM. Fig. \ref{fig:in_ex} shows the distribution of the original locations of all 2G star progenitors which finished within a radius of 20 pc from the centre of the 1G in the final time step. The navy ex-situ curve and gold in-situ curves are scaled according to their total contribution to the 2G. With respect to the centre of the parent galaxy, the maximum value of the probability density distribution demonstrates that the majority of 2G material originated from $\approx$ 500 pc from the galaxies centre. This is reasonable given the GCs initial offset from the centre and radius of the HI hole, as seen in Fig. \ref{fig:orbit time ev}. This figure shows that there is a significant spread in the initial positions of the gas accreted by the GC, with the maximum value of an ex-situ particle which made its way into the GC at 1750 pc. This introduces a potential problem for our scenario; gas particles from such a large radii may introduce a non-negligible [Fe/H] spread which is not detected in Galactic GCs. However, as we are analysing the starting positions of all new stars within a 20pc radius, Fig. \ref{fig:he} and \ref{fig:in_ex_situ} demonstrate that it is questionable whether anything at radii larger than 10 pc is a genuine 2G star. Future investigations into the long term evolution will again affect this distribution and whether this spread in initial locations of 2G stars is a major issue for this model.

\subsection{Scaling Relations}
The results given in \cite{milone_etal2020} demonstrate the variety of $f_{2G}$ vs GC mass relations which exist for GCs and how their parent galaxy can influence the gradient of the slope. Here we present similar scaling relations, derived from our simulations. The results should be taken with caution as the dynamical effects such as stripping and SNe have the potential to dramatically influence the correlation between the 2G and total mass of a cluster. Here we summarise the most notable correlations using a data set of simulations which is representative of Galactic GCs. 34 clusters meet the criteria which include having a realistically sized HI hole, thus allowing gas accretion to occur within 370 Myr. For the present simulations, this translated into a radius of no more than 300 pc. We also require that the 1G is not destroyed in some way by the parent galaxy. This results in setting a minimum size threshold on possible progenitors. Several simulations ran for only 150 Myr but have been excluded as the proto-GCs are usually in a far more turbulent state, and for some models, have not completely finished their accretion process. We also provide a sample of AGB only simulations which we achieve by modifying the HI hole to be larger than 500 pc, artificially inhibiting any gas accretion, which we use as a comparison for our accreting GC models. Finally we present a sample of GCs formed in younger galaxies or low surface brightness galaxies (LSB), both of which are gas poor when compared to our fiducial model's galaxy. The parameters for each of these models are available in Tables \ref{table:all_sims}, \ref{table:HI_sims} and \ref{table:LSB_sims} in the Appendix.

\subsubsection{The $f_{enriched}$ vs $M_{GC}$ relation}

\begin{figure}
	\includegraphics[width=\columnwidth]{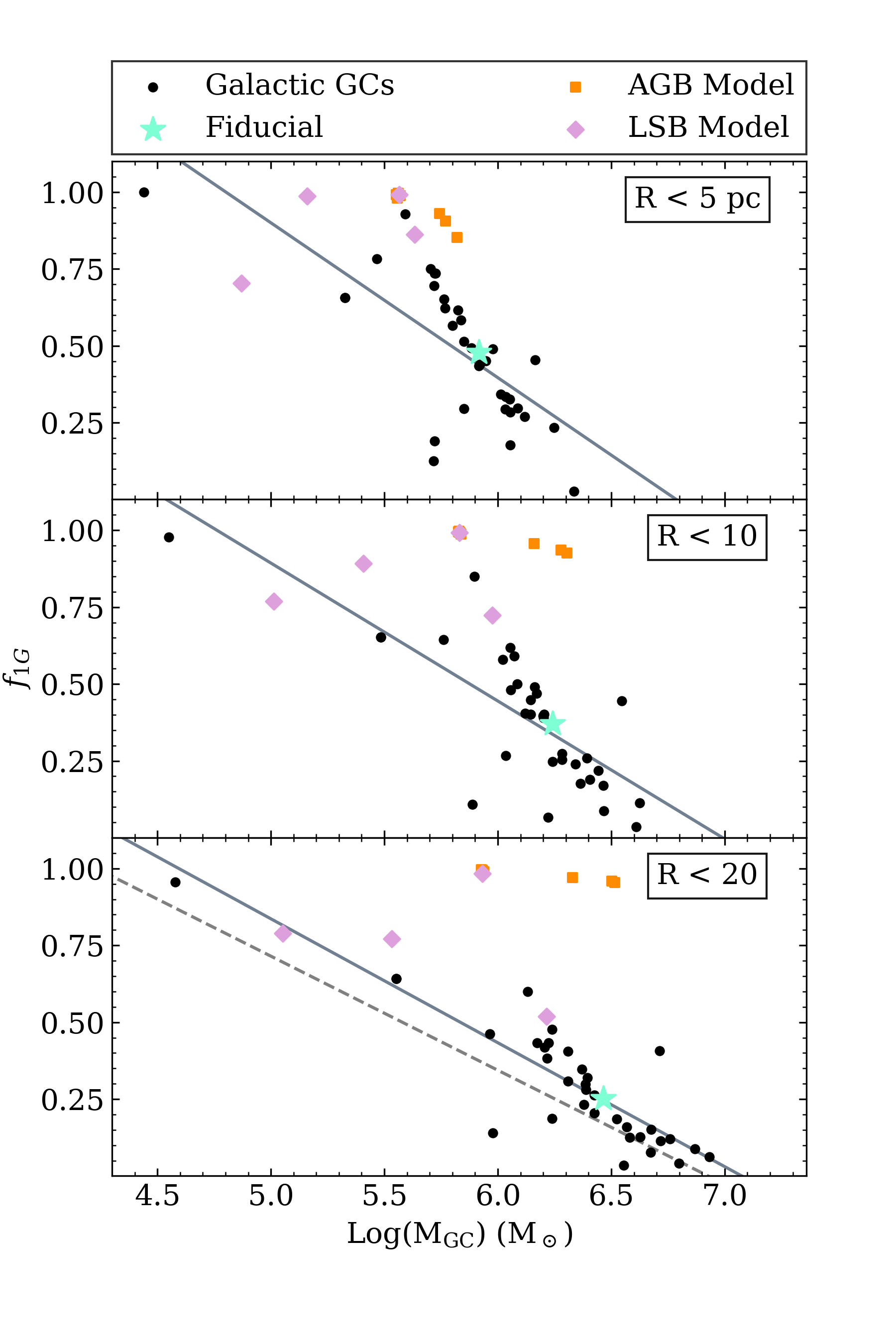}
    \caption{$f_{1G}$ as a function of GC mass within a 5 pc, 10 pc and 20 pc radius. The line of best fit, shown by a solid grey line, is calculated using only the Galactic GC data. The equations of these lines can be found in Table. \ref{tab:correlations}. The dashed line in the bottom panel is representative of the initial masses predicted by \citealt{Baumgardt_etal2019}, taken from Fig. 6 in \citealt{Gratton_etal2019}.}
    \label{fig:enriched_fraction}
\end{figure}

The relationship between the mass of the 2G and the total mass of the cluster is one of the most notable results arising from the last decade of GC research. From our simulations we recover a correlation which agrees with the results from observations. To be consistent with prior works in the literature, in Fig. \ref{fig:enriched_fraction} we plot this relationship relative to the fraction of 1G stars within the cluster, $f_{1G}$. We expect to find small clusters dominated by 1G stars residing in the top left, and more massive 2G dominated clusters in the bottom right. The values represent the initial masses of possible Galactic GC progenitors immediately after the 2G has been formed. For the Galactic GC models, the 1G half mass radius was approximately at 5 pc. We do not consider the radius of the 2G as the stars tend to be in a turbulent state. For all stars within a 5 pc cut off, there is a moderate anti-correlation between the total mass of the cluster and fraction of 1G stars (see Table \ref{tab:correlations}). A steeper gradient may exist for clusters within the mass range of $10^{6.5} \rm{M}_{\odot}$ to $10^{5.5}\rm{M}_{\odot}$, however outlier points on the lower mass end mask this result. Increasing the cut off radius to 10 pc, which in most cases contains the majority of 1G mass (see Fig. \ref{fig:f_enriched}), this alternative steeper gradient still exists but the correlation between the mass of the cluster and the fraction of 1G stars improves. Finally, we see our strongest correlation at a radius of 20 pc. A useful comparison for these values is initial masses of the Galactic GCs presented in \cite{Baumgardt_etal2019}. The data clusters at a slightly higher GC mass than the results predicted by \cite{Baumgardt_etal2019}; however, we expect this is more of a selection effect of our study primarily focusing on large, 47 Tuc-like analogues. Our fiducial model discussed in the previous section lies on top of the predicted curve. Previous investigations into GC formation scenarios have emphasised that pristine gas accretion is necessary in explaining the O-Na anticorrelation. Our study adds to this argument that gas accretion is required in order to reproduce the relationship between the initial mass of the cluster and the fraction of 2G stars. However, this result relies on the assumption that all 2G stars within a 20 pc radius are accreted by the GC. Fig \ref{fig:time_ev_final_GC} and \ref{fig:f_enriched} illustrates that at least for the fiducial model, the outer regions of 2G stars may be preferentially stripped from the cluster during further evolution. As this is a determining factor in the success of this model, we intend to study the long term evolution of the cluster in more detail in future works. Our model emulates Magellanic Cloud GCs or GCs which originated in younger galaxies by simulating Low Surface Brightness (LSB) galaxies. The clusters from these simulations always appear more 1G dominated than their Galactic counterparts and lie towards the right of the line of best fit. This agrees with the results presented in \cite{milone_etal2020} that Magellanic cloud clusters have a lower fraction of 2G stars than Galactic GCs.

Fig. \ref{fig:enriched_fraction} also illustrates the lower limit on the mass of the 1G in order for a realistic 2G to form. M21 in Table \ref{table:all_sims} represents the lowest mass GC that was able to retain a 2G population. As the size of the GC decreases, it starts to become more susceptible to the gravitational influence of larger clusters which formed within the Galaxy. However, when in a LSB galaxy, the lower limit for 2G formation differs as there are fewer massive clusters which disturb the gas around the 1G. This can be seen for M44 in Table  \ref{table:LSB_sims}, particularly within the central 5 pc of the cluster. Here we would predict a much lower 2G fraction for that mass, however, altering the parameters of the host galaxy allows this model to account for a wide range of observations.

\begin{figure}
	\includegraphics[width=\columnwidth]{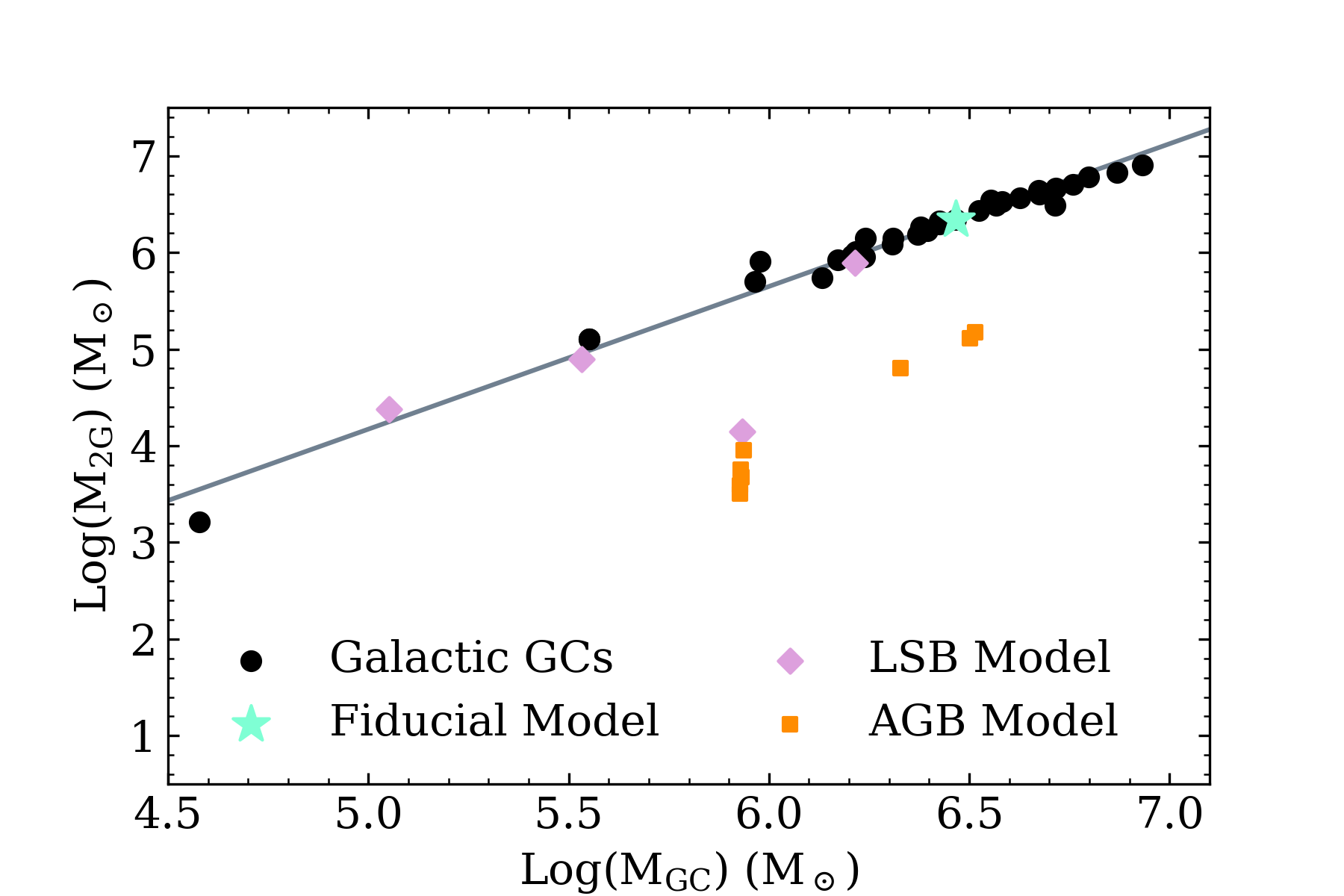}
    \caption{The mass of the 2G against the total mass of the cluster within 20 pc. The Galactic GC points are tightly correlated. AGB only models do not follow this trend. The coefficients extracted from fitting lines to these data sets can be found in Table \ref{tab:correlations}.}
    \label{fig:2G_GC}
\end{figure}

We exclude Bondi accretion \citep{Bondi_1952} as the main mechanism by which gas is accreted onto the GC. To confirm this, we plot the mass of the 2G against the mass of our Galactic GCs in Fig. \ref{fig:2G_GC}. The coefficients listed in Table \ref{tab:correlations} demonstrates a strong correlation between the 2G and total GC mass and that the corresponding gradient is approximately 1.5. Bondi accreation assumes that the amount of accreted material is proportional to the square of the stellar mass. Our derived slope is not as steep as that predicted by spherically symmetrical accretion, and we assume that gas accretion is due to the gravitational potential of the 1G. In comparison to this, Table \ref{tab:correlations} also includes the results from fitting our AGB only models. Although there is a much smaller sample size, the line of best fit has a gradient of 2.55. We expect this to be different to our Galactic GC models as no gas accretion is involved.

\subsubsection{The rotation amplitude vs $M_{GC}$ relation}
\label{sec:rot_amp}

\begin{figure}
	\includegraphics[width=\columnwidth]{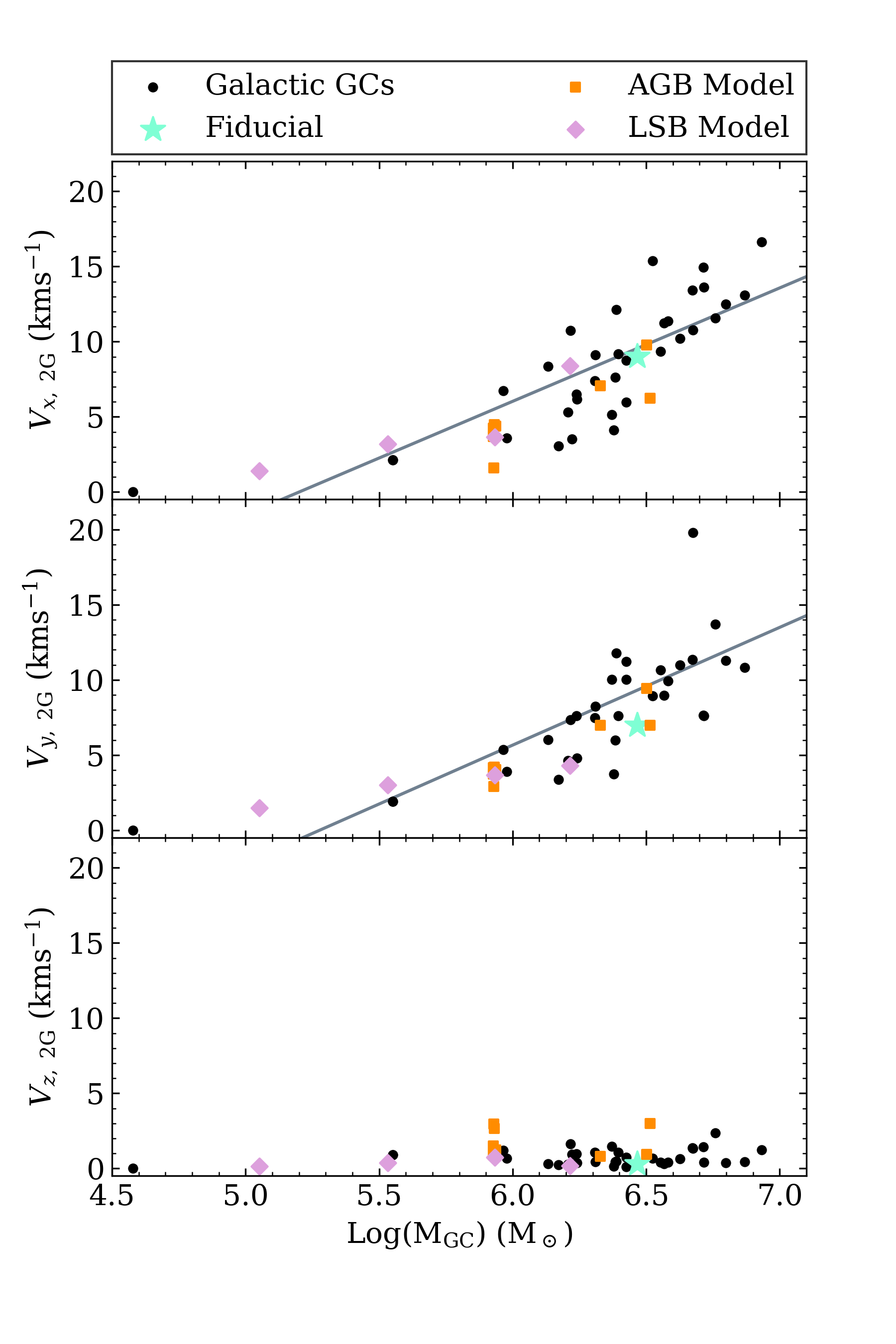}
    \caption{Final mass of the GC as a function of the rotation amplitude in the X, Y and Z directions. The coefficients for the lines of best fit are found in Table. \ref{tab:correlations}. }
    \label{fig:kinematics_scale}
\end{figure}

One unexpected correlation found during the development of this scenario was the connection between the final mass of the cluster and the rotation amplitude of 2G stars. By extracting the amplitude from both populations (see Fig. \ref{fig:polar_rot}), Fig \ref{fig:kinematics_scale} compiles the results from our physical simulation samples to illustrate that the rotation of the 2G is predicted to be higher for more massive clusters. Table \ref{tab:correlations} lists a moderate positive correlation for $V_x$ and $V_y$ vs $\rm{log_{10}(M}_{GC})$ parameter pairs. No correlation exists for the $V_z$ - $\rm{log_{10}(M}_{GC})$ parameter pair as this represents the axis of rotation of both the cluster and the parent galaxy. Similarly, the difference between the maximum rotation amplitude of the 1G and 2G exhibit a moderate positive correlation. We expect this result to be invariant to the host galaxy parameters as this trend is also evident in our LSB models. This finding conflicts with current observations of Galactic GCs which, for the most part, show little to no difference between the rotation amplitudes of the two populations (e.g. see \citealt{Cordoni_etal2020}). Furthermore, this relation implies that larger clusters will exhibit signs of internal rotation. Current studies of internal kinematics using Gaia DR2 have shown this not to be the case, although future data releases with more precise measurements may alter this view. It is unclear how the dynamical evolution of our simulated GCs will progress given the different rotation characteristics. We hope that the long-term evolution can significantly weaken the rotation amplitudes of the two populations and we intend on investigating this question in future studies

Another kinematic prediction is that even when the initial internal axis of rotation ($\theta_{\rm{gc}}$) for the 1G is altered, the accretion processes forces the rotation axis of the proto-GC to align with the parent galaxy. For example, M27 (see Table \ref{table:all_sims}) initially starts rotating at an angle of $60^\circ$ with respect to its host galaxy and after the 2G formation process, rotates in line with the galaxy. The corresponding AGB only simulation, M40, retains this initial internal rotation for both the 1G and 2G. In most cases, the 2G in AGB only models share a similar phase and amplitude to the 1G. This is an important conclusion for our scenario as it suggests that the host galaxy leaves a kinematic imprint on the cluster. We discuss this point further in Section \ref{sec:internal_spin}.

\begin{table}
\centering
	\caption{Correlations seen between different sets of parameters for the Galactic GC sample. Correlations are determined by fitting the line $y = ax + b$ to the corresponding data sets. Unless specified, the clusters were set to an arbitrary 20 pc cut off. ($^*$ corresponds to our AGB only model.)}
	\label{tab:correlations}
    \begin{tabular}{lllll}
    \hline
Parameter Pair  & a      & b       & r & p-value \\
    \hline
$\rm{f}_{1G}$ - $\rm{log}_{10}(\rm{M}_{GC})$ < 5 pc    & -0.504 & 3.418   & -0.729      & 6.90e-07 \\
$\rm{f}_{1G}$ -  $\rm{log}_{10}(\rm{M}_{GC})$ < 10 pc  & -0.449 & 3.137   & -0.768      & 7.02e-08\\
$\rm{f}_{1G}$ -  $\rm{log}_{10}(\rm{M}_{GC})$          & -0.403 & 2.851   & -0.868      & 1.47e-11 \\
    \hline
$\rm{log}_{10}(\rm{M}_{2G})$ -  $\rm{log}_{10}(\rm{M}_{GC})$  & 1.477  & -3.211  & 0.987 & 4.08e-27\\
$\rm{log}_{10}(\rm{M}_{2G})$ -  $\rm{log}_{10}(\rm{M}_{GC})^*$  & 2.553  & -11.442  & 0.984 & 1.05e-05\\
    \hline
$V_{x, 2G}$ -  $\rm{log}_{10}(\rm{M}_{GC})$     & 7.544  & -39.236 & 0.798  & 9.50e-09\\
$V_{y, 2G}$ -  $\rm{log}_{10}(\rm{M}_{GC})$     & 7.834  & -41.342 & 0.713  & 1.55e-06\\
$V_{z, 2G}$ -  $\rm{log}_{10}(\rm{M}_{GC})$     & 0.271  & -0.966  & 0.229  & 1.87e-01\\
    \hline
$V_{x, 2G - 1G}$ -  $\rm{log}_{10}(\rm{M}_{GC})$ & 9.001  & -45.941 & 0.800  & 8.39e-09\\
$V_{y, 2G - 1G}$ -  $\rm{log}_{10}(\rm{M}_{GC})$ & 7.996  & -42.379 & 0.713  & 1.56e-06\\
$V_{z, 2G - 1G}$ -  $\rm{log}_{10}(\rm{M}_{GC})$ & 0.433  & -2.003  & 0.338  & 4.73e-02\\       
    \hline
    \end{tabular}
\end{table}

\section{Discussion}
\label{discussion}

\subsection{The origin of the $\rm{M}_{2G}$ vs $\rm{M}_{GC}$ relation and the link to LSB galaxies}
\label{sec:2G_origin}

The results from Fig. \ref{fig:enriched_fraction} and \ref{fig:2G_GC} illustrate how the properties of the parent galaxy can influence the fraction of 2G stars within the GC. The gas fraction of the galaxy is the primary factor which mediates the fraction of 2G stars. We also find a lower limit on the initial mass of a cluster in which we can establish a 2G population. Clusters which are born closer to the galactic centre experience more clumpy accretion events due to their densely populated environment. This results in larger 2G populations for massive clusters, however there is a higher chance that a cluster may be tidally destroyed by its siblings. Conversely, clusters which do not pass through the galactic centre and born at larger radii have a high fraction of in-situ star formation within the 2G population and are less likely to be tidally disrupted. This enables low mass GCs to survive and generate a feasible 2G of stars. The metallicity of the galaxy has minimal consequences on the 2G fraction or morphology of the resulting cluster, however it does impact the He enrichment, with higher metallicity galaxies generating slightly more He-rich 2Gs. Modelling the chemistry of the clusters is difficult with the present simulations and thus we do not dwell on these results.

As we suggest that the gas density of the host galaxy is one of the primary drivers of the fraction of 2G stars, we expect that GC formation is not necessarily a product of the early universe. Observations have demonstrated that high redshift galaxies are significantly more gas-rich compared to z = 0 galaxies (e.g. \citealt{Daddi_etal2010}). Young stellar clusters may also be able to develop GC like qualities provided that they are exposed to extremely high gas fractions as were common in the early universe. GCs from LMC-like and other low surface brightness galaxies tend to have a lower fraction of 2G stars when compared to the Galactic counterparts. As found by \cite{milone_etal2020}, LMC clusters are more 1G dominated and do not lie on the line of best fit for Galactic GCs and we see this same phenomenon occurring in our models. A lower gas density can have the effect of lowering the minimum mass that is needed in order to create a 2G as illustrated by M44 from Table \ref{table:LSB_sims}.

\subsection{The chemical implications of clumpy accretion}
\label{sec:clumpy_discussion}

The phenomenon of clumpy accretion allows our model to explain several observable parameters, including the expected mass of 2G stars and a connection to the parent galaxy's parameters. However, it introduces several unknown factors concerning the chemical composition of the cluster. Given the vast range of radii that the accreted gas particles originated from in Fig. \ref{fig:in_ex}, this accretion process may induce unwanted chemical abundance spreads. Observationally, it is challenging to estimate the [Fe/H] gradient for high-z GC-hosting dwarf galaxies. Taking the LMC as a proxy, the [Fe/H] gradient is shallow (-0.01 dex/kpc for old stars, Fig.2 from \citealt{Feast_etal2010}). Thus if a GC forms from gas spreading over a 1 kpc region, then  Fe/H spread is only 0.01 dex for the 2G alone. An analysis of globular clusters from the APOGEE survey by \cite{meszaros_etal2020} stated that the true iron spread for their GCs was on average 0.068 dex across their 30 clusters. For our fiducial model, gas accreted to form the 2G spans a radius of 1.75 kpc and depending on the slope of the [Fe/H] gradient employed for our galaxy, the final spread in metallicity could be within the range set by observational results.

Clumpy accretion comes with many more obstacles than inducing a spread in [Fe/H]. By delaying the accretion of 2G clumps, this scenario can naturally explain Type II clusters such as NGC 2808. However, if clumpy accretion is to occur for Type I GCs, it would need to take place in the very early stages of the 2G formation process. Even still, this may induce changes in the chemical makeup of the 2G. The merging of GCs is discussed in \cite{vandenBergh_1996} as a mechanism for producing composite color-magnitude diagrams. Mergers may have occurred in dwarf spheroidal galaxies with low internal velocity dispersion. At T = 0 of the fiducial model, the 1D velocity dispersion of disc stars is $19 kms^{-1}$. This low dispersion may relate to the incidence of clumpy accretion.

High precision measurements of Galactic GCs have uncovered slight variations in populations within some regions of a cluster's colour-magnitude diagram (CMD). For example, \cite{DiCriscienzo_etal2010} noted that 47 Tuc is comprised of three different subpopulations; a first-generation consisting of 30\% of the stars in the cluster, a SGI (equivalent to a 2GI in our nomenclature) which makes up 60\% of stars belonging to the second generation and lastly a SGII (2GII nomenclature) which makes up 10\% of 2G stars and is found within the faint subgiant branch in the CMD. Our model can rationalise the existence of these 2G subpopulations by attributing them to slight differences in the location these populations were formed. More enriched populations could originate within the centre of the GCs where the gas is heavily enriched with He from AGB ejecta, whereas less He rich stars could form in similar conditions within the galaxy and then be later accreted by the cluster. Another example of this phenomenon documented by \cite{milone_etal2020} is found in the Small Magellanic Cloud (SMC) cluster NGC121 which also has two enriched components. Clumpy accretion could still be possible for low surface brightness galaxies such as the SMC and thus our scenario could potentially explain this observation. This is all, of course, speculative and requires both even higher resolution simulations and more precise observations with which we can compare our results.

\subsection{2G concentration and long term evolution}
\label{sec:2G_concentration}
Fig. \ref{fig:f_enriched} illustrates that the outskirts of the cluster are entirely 2G dominated. This issue plagues the majority simulations which showed external gas accretion. In its current state, the distribution of 2G stars around the cluster poses issues for this scenario as observations show that 2G stars are generally more centrally concentrated \citep{Lardo_etal2011}. The 2G stars form in a disc, parallel to the plane of the galaxy rather than showing spherical symmetry like the 1G. To reproduce observations, the cluster must avoid having its 2G stars being stripped by the galaxy and redistribute them towards the centre. However, the Type II clusters M15 and M80 both show a central concentration of 1G stars (\citealt{Larson_etal2015}; \citealt{Dalessandro_etal2018M80}) and thus clumpy accretion may be the cause of these observations. The cores of GCs with a higher central concentration of 2G stars tend to be more ex-situ star dominated. As ex-situ stars tend to be less He rich compared to their AGB gas enriched in-situ counterparts, this would influence any chemical gradients within the cluster. On the assumption that all 2G stars within a 20 pc radius of the 1G mass are retained by the GC and that the 2G stars become more centrally concentrated as the cluster relaxes, this would allow this scenario to solve the mass budget problem. This result agrees with a study of pristine gas accretion by \cite{Calura_etal2019}, who determined that the degree of 1G mass loss required to produce observed ratios of 1G to 2G in GCs is less extreme than what the mass budget problem suggests.

Our fiducial model overestimate the mass of GC progenitor to allow for some degree of mass loss over its life time. Additionally, there will be a further reduction of 2G mass due to future SNe events. However, these events will be necessary to remove the non-negligible amount of gas which remains within the proto-cluster. For our fiducial model, the simulation ends with $\approx 6\times10^4 \rm{M}_{\odot}$ of gas within 20 pc of the centre of mass. This gas must be expelled in order to explain the lack of neutral ISM observed in Galactic GCs. Simulations by \cite{Chantereau_etal2020} found that both ram pressure stripping and ionisation is mandatory to explain the small amount of ionized gas in the core of GCs. We intend to investigate these channels of gas expulsion in future simulations of our Galactic GCs.

\subsection{The physicality of a large HI hole}
\label{sec:Large_HI}

An evacuated region of gas is placed around the progenitor cluster to account for SNe effects from the 1G on the surrounding ISM. This HI hole acts to suppress the early accretion of gas onto the cluster. We admit that this is a gross simplification, however, instances of HI holes appearing in dwarf galaxies have been recorded in the literature. \cite{Warren_etal2011} discussed possible origins of kpc-scale holes in the atomic hydrogen distributions of some nearby dwarf irregular galaxies. Using radial analysis, they calculate the HI hole size for DDO 181 to be $\approx740$ pc, Holmberg I to be $\approx850$ pc and M81 Dwarf A to be $\approx745$ pc as three examples. Each of these radii fit comfortably in the range we set for our cluster progenitors. However, different methods can result in vastly different predictions of hole size. Using the THINGS \citep{Walter_etal2008} survey, \cite{Bagetakos_etal2011} identified over 1000 HI holes in their sample, and as an example, found six holes in Holmberg I, ranging in size from 190 pc to 740 pc. \citet{Vorobyov_etal2004} used numerical simulations to investigate the energy required to create such a hole in Holmberg I. Through their modelling they deduced that a hole of this magnitude required an energy equivalent to $300 \pm 50$ Type II supernovae. The energy our 1G is capable of expelling will be tightly linked to its IMF. A back of the envelop calculation assuming a 1G population with a Salpeter IMF of $\alpha = 2.35$ and mass range of 0.1 - 50 $\rm{M}_{\odot}$ would require $7 \times 10^{-3} \rm{M}_{\odot}$ SNII per $\rm{M}_{\odot}$. We assume $4.2 \times 10^{4} \rm{M}_{\odot}$ for HI hole formation and thus with our $10^6 \rm{M}_{\odot}$ fiducial cluster, a 200 pc hole could be assumed to be a conservative estimate. Although the radius of the HI hole ($R_{\rm{HI}}$) is treated as a free parameter in our model, it is strongly influenced by feedback effects of the 1G. This in turn impacts the duration of the holes collapse and thus the subsequent gas accretion onto the GC. We intend to discuss this further during investigations into 1G formation and HI hole creation in order to rationalise our present assumptions.

Models which do not include HI holes overestimate the total mass of the cluster. Secondly, there is no delay in star formation and therefore, no distinction between 1G and 2G stars. Gas that would have been consumed during 1G formation is readily available to the cluster to immediately start forming the 2G. There should be some physical mechanism which prevents this from occurring and thus some form of HI hole is necessary in reproducing observations.

\subsection{Spin axis alignment between the GC spin axes and the parent dwarf}
\label{sec:internal_spin}

As discussed in Section \ref{sec:kinematics}, the accretion process causes the internal rotation axis of the GC to align with that of the parent galaxy, independent of the GCs host environment. This prediction has implications for the identification of accreted clusters and can potentially aid in reconstructing the accretion history of the Milky Way (MW). Taking the 'Sausage' GCs identified by \cite{Myeong_etal2018SausageGC} as an example, our scenario would predict that the rotation axis of the integrated orbits of these clusters would align with the internal rotation of the Gaia-Enceladus dwarf before it was destroyed by the MW. Additionally, this method could contribute to the authentication of any potential Sausage clusters. We note that this may only be used for rotating clusters. NGC 2808, NGC 7089 (M2) and NGC 1904 (M79) are three Gaia Sausage clusters as identified by \cite{Myeong_etal2018SausageGC} and have evidence of rotation as found by \cite{sollima_etal2019}. By integrating the orbits of these clusters and analysing their axis of rotation at the time they were accreted, they could be used to test our hypothesis. This is reliant on the long term evolution not erasing the kinematic signature from the host. 

\subsection{Disc stars as an additional metal poor population}
\label{sec:disc stars}
As evident in Fig. \ref{fig:mass time ev}, the simulation starts with $10^5 \rm{M}_{\odot}$ of disc stars surrounding the 1G. The concentration of these stars is visible in the left most panel of Fig. \ref{fig:final_ts} and as the system evolves this mass remains relatively constant. Fig. \ref{fig:time_ev_gal} demonstrates that the clustering of disc stars also occurs for several of the fiducial model's sibling clusters. This phenomenon is found across the whole data set, and is not dependent on any free parameters. From this we conclude that GCs should be capable of retaining a small population of disc stars from their parent galaxy. The $\sigma_{1D}$ velocity dispersion of the disc stars within a 50 pc radius of the cluster is $15 kms^{-1}$. Our fiducial GC with a mass of $~3\times10^6 \rm{M}_{\odot}$ and radius of 20 pc could potentially capture these disc stars as the internal dispersion of the cluster is $\sqrt{\frac{GM_{\odot}}{R}} \approx 25 kms^{-1}$. This may explain the origin of the 3rd metal poor population seen on the subgiant branch in 47 Tuc (\citealt{milone_etal2012}), or the small 3rd generation seen in Terzan 5 (\citealt{Ferraro_etal2009}). These captured disc stars are a feature in all simulations where the 1G progenitor was born in a central region of the galaxy. Our simulations predict that most Galactic GCs should comprise of a metal-poor component with the percentage of total mass varying depending on the dynamical evolution of the cluster. Future works intend on taking a much deeper look into the formation processes of how a 1G may capture these stars in the first place and how these stars may evolve after the cluster has been formed.

\section{Conclusions}
\label{conclusion}
We have presented the preliminary results from our original, hydrodynamical simulations which builds upon the theoretical framework first proposed by \cite{DErcole_etal2008}. In our model, we have first identified GC progenitor gas clouds with masses of $10^7 \rm{M}_{\odot}$. Next, these clouds are converted into compact stellar systems associated with the 1G. Around this system, we included a HI hole produced by the multiple 1G SNe to demonstrate how SNe can influence the gas accretion processes and subsequent star formation. The gas accretion process proceeds after the collapse of this hole and ISM from the galaxy mixes with AGB ejecta from 1G stars. This resulted in the birth of a secondary population of stars known as the 2G.

The foremost result from our study is that we have been successful in producing a correlation between the mass of the cluster and the fraction of 1G stars. Additionally, the gas fraction of the galaxy is the dominant parameter in mediating this relationship with high gas fraction galaxies producing GCs which are more 2G dominated. The gas fraction also sets the lower limit for which no 2G is observed due to gas poor galaxies within the simulation experiencing less turbulence which can lead to the disruption of the GC. Prior to the formation of the 1G, we find that the GMC has the potential to capture a population of disc stars. We predict that these stars are the missing minor, metal-poor component that has been observed in some Galactic GCs. Additionally, we expect that a very small population should be present in the majority of GCs. We intend to examine this prediction further in our next paper.

We observe GC progenitors interacting and accreting one another through a process we have termed clumpy accretion. Clumpy accretion is a hierarchical process and can occur at any point during the simulation. However, we suggest that bona fide Type I clusters can only undergo clumpy accretion within the first 50 Myr as to minimise the risk of inducing a [Fe/H] spread. We argue that this form of accretion may be beneficial in producing smaller subpopulations of 2G stars observed in Galactic GCs. Type II clusters on the other hand, may experience clumpy accretion at any point within the 370 Myr window of the simulation as this would give rise to distinct populations characterised by multiple [Fe/H] values. Our scenario suggests that Type I and Type II GCs share the same origin. 

The kinematic repercussion of our scenario is that the 2G of a GC exhibits highly coherent rotation in a disc-like structure, parallel to the plane of the parent galaxy. As the long term dynamical effects of this rotation is unclear, we intend to perform follow up simulations on our cluster progenitors in future studies. If the GC retains information about the plane of the host galaxies rotation, this information could be used to inform cosmological simulations about the accretion history of the Milky Way. Provided this kinematic imprint of the parent galaxy remains within the cluster, we predict that by backwards integrating the motion of confirmed ex-situ GCs and analysing their axis of rotation, it may be possible to infer the plane of the dwarf galaxy prior to its accretion.

The cornerstone of this model is its ability to reproduce the observed correlation between the GC's mass and its fraction of enriched stars. However, this result relies on the accretion of all 2G stars within a 20 pc radius of the GC. Further long-term dynamical investigations are required to determine the outcome of these stars as this could lead to the falsification of our results. In its current state, this scenario is unable to explain all observational criteria for Galactic GCs. Nevertheless, it still holds scientific merit as being the first of its kind to simultaneously model the dynamics of MSP formation within a GC and the entirety of its parent galaxy. We intend to continue to improve and develop this model to more closely align with the observed characteristics of GCs. Gaia DR3 will shed more light on the internal kinematics of GCs while spectroscopy from instruments such as VLT/MAVIS and photometry from the James Webb Space Telescope will better quantify the chemistry of the populations. Further developments in AGB yields will allow us to refine the degree of mixing with pristine gas to constrain our parameters further. We have paved the way for self-consistent simulations of GC formation scenarios and look forward to comparing our results with future groups who undertake similar investigations.

\section*{Acknowledgements}
The authors thank the anonymous referee for their kind and thoughtful comments. Their suggestions helped to identify ambiguities and improved the clarity of this work. Numerical simulations were run
on the Pleiades and OzSTAR GPU clusters kindly made available to
us through the International Center for Radio Astronomy Research at
The University of Western Australia and the Centre for Astrophysics
and Supercomputing at the Swinburne University of Technology. This study made use of the \textsc{PYTHON} packages \textsc{NUMPY} \citep{van2011numpy}, \textsc{SCIPY} \citep{2020SciPy-NMeth}, \textsc{MATPLOTLIB} \citep{matplotlibHunter:2007}, \textsc{PANDAS} \citep{PANDAS} and \textsc{IPYTHON} \citep{ipythonPER-GRA:2007}.

\section*{Data Availability}
The data underlying this article will be shared on reasonable request to the corresponding author.




\bibliographystyle{mnras}
\bibliography{Main} 



\appendix
\section{Spatial distributions for the remaining components within the simulation}
\label{sec:spatial_comp_app}
The companion figures of Fig \ref{fig:time_ev_new} for the gas and disc star components.
\begin{figure*}
    \label{fig:time_ev_gas}
	\includegraphics[width=0.9\textwidth]{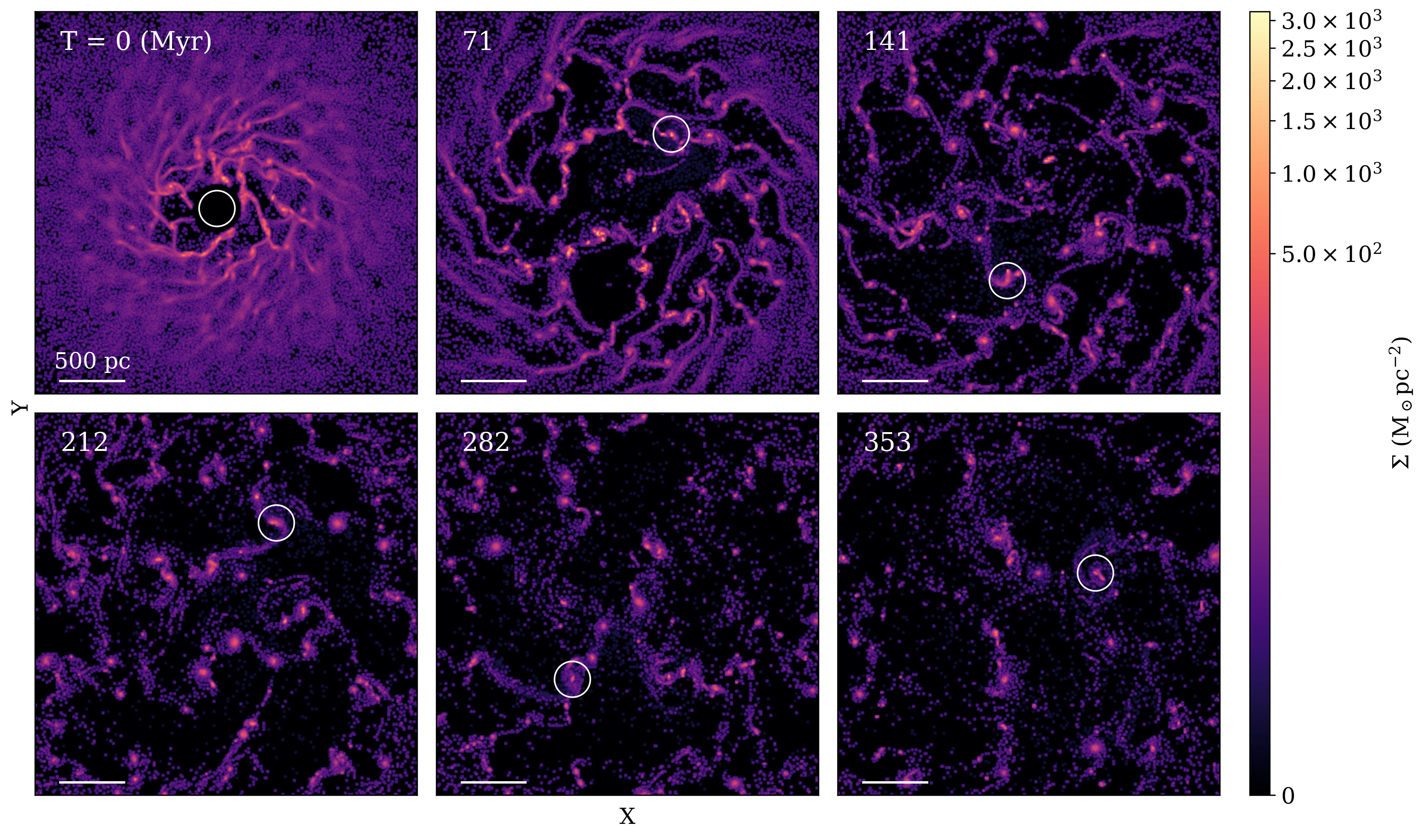}
    \caption{The spatial distribution of gas particles during the simulation. This includes both ISM and AGB ejecta. The figure shares the same properties as Fig. \ref{fig:time_ev_new}.
    }
\end{figure*}

\begin{figure*}
    \label{fig:time_ev_gal}
	\includegraphics[width=0.9\textwidth]{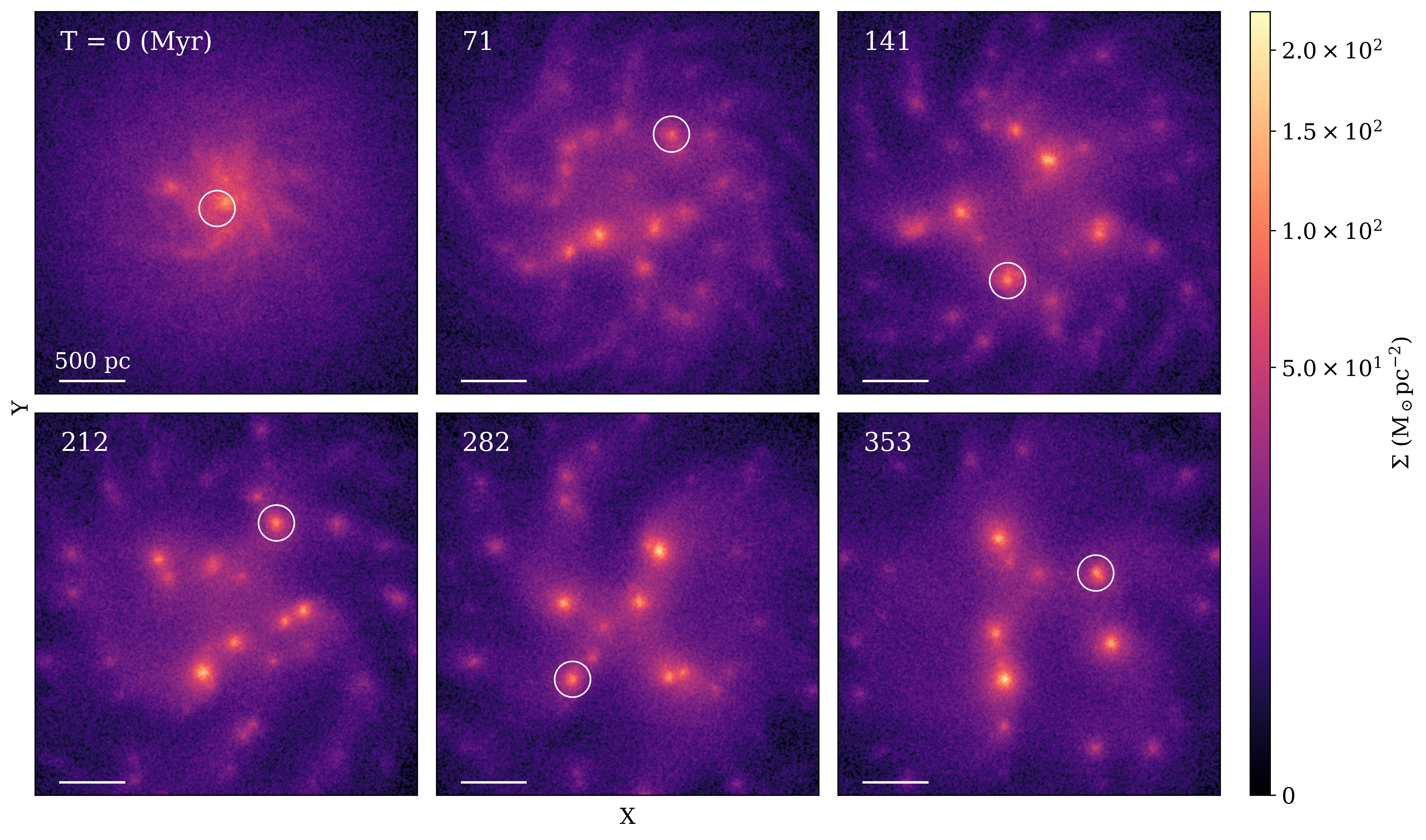}
    \caption{The spatial distribution of disc stars during the simulation. The figure shares the same properties as Fig. \ref{fig:time_ev_new}.
    }
\end{figure*}

\section{Simulation Models}
\label{sec:appendix_tables}
We present the results of all 370 Myr simulations which were used to construct the scaling relation in Fig \ref{fig:enriched_fraction}. We do not include our larger sample of 140 Myr simulations which were used to inform our current parameter choices.

\begin{table*}
    \centering
	\caption{Parameters for our Galactic GC models which ran for 370 Myr. $\rm{R}_{start}$ denotes the distance of the GCs starting location with respect to the host galaxy's centre of mass.}
    \begin{tabular}{lcccccccr}
    \hline
    Model ID & $M_{\rm{dm}}$ ($\rm{M}_{\odot}$) & $\rm{R}_{galaxy}$ (kpc) & $M_{gc}$ ($\rm{M}_{\odot}$) & $\rm{R}_{gc}$ (pc)& $\rm{R}_{HI \ hole}$ (kpc)& $f_{\rm rot}$ & $\theta_{\rm gc}$ ($^{\circ}$)& $\rm{R}_{start}$ (pc) \\ 

    \hline
    M1 & 5$\times 10^9$ & 2.4 & 1$\times 10^6$ & 0.02 & 0.2 & 0.2 & 0 & 85 \\ 
    M2 & 5$\times 10^9$ & 1.7 & 1$\times 10^6$ & 0.02 & 0.1 & 0.2 & 30 & 36 \\ 
    M3 & 5$\times 10^9$ & 1.7 & 1$\times 10^6$ & 0.02 & 0.1 & 0.2 & 30 & 187 \\ 
    M4 & 5$\times 10^9$ & 1.7 & 1$\times 10^6$ & 0.02 & 0.3 & 0.2 & 30 & 36 \\ 
    M5 & 5$\times 10^9$ & 1.7 & 1$\times 10^6$ & 0.02 & 0.3 & 0.2 & 30 & 36 \\ 
    M6 & 5$\times 10^9$ & 1.7 & 1$\times 10^6$ & 0.02 & 0.2 & 0.2 & 30 & 179 \\ 
    M7 & 5$\times 10^9$ & 1.7 & 3$\times 10^5$ & 0.02 & 0.3 & 0.2 & 30 & 36 \\ 
    M8 & 5$\times 10^9$ & 1.7 & 3$\times 10^5$ & 0.02 & 0.2 & 0.2 & 30 & 36 \\ 
    M9 & 5$\times 10^9$ & 1.7 & 1$\times 10^6$ & 0.02 & 0.2 & 0.2 & 30 & 38 \\ 
    M10 & 5$\times 10^9$ & 1.7 & 1$\times 10^6$ & 0.02 & 0.2 & 0.2 & 0 & 130 \\ 
    M11 & 5$\times 10^9$ & 1.7 & 1$\times 10^6$ & 0.02 & 0.2 & 0.2 & 0 & 370 \\ 
    M12 & 5$\times 10^9$ & 1.7 & 3$\times 10^5$ & 0.02 & 0.2 & 0.2 & 30 & 36 \\ 
    M13 & 5$\times 10^9$ & 1.7 & 1$\times 10^6$ & 0.02 & 0.2 & 0.2 & 0 & 83 \\ 
    M14 & 5$\times 10^9$ & 1.7 & 1$\times 10^6$ & 0.02 & 0.2 & 0.2 & 0 & 466 \\ 
    M15 & 5$\times 10^9$ & 2.4 & 1$\times 10^6$ & 0.02 & 0.1 & 0.2 & 0 & 85 \\ 
    M16 & 5$\times 10^9$ & 2.4 & 1$\times 10^6$ & 0.02 & 0.2 & 0.2 & 0 & 450 \\ 
    M17 & 2$\times 10^{10}$ & 3.4 & 1$\times 10^6$ & 0.02 & 0.2 & 0.2 & 0 & 140 \\ 
    M18 & 5$\times 10^9$ & 1.7 & 1$\times 10^6$ & 0.02 & 0.2 & 0.2 & 0 & 152 \\ 
    M19 & 2$\times 10^{10}$ & 4.9 & 1$\times 10^6$ & 0.02 & 0.2 & 0.2 & 0 & 152 \\ 
    M20 & 5$\times 10^9$ & 2.4 & 1$\times 10^6$ & 0.02 & 0.2 & 0.2 & 0 & 18 \\ 
    M21 & 5$\times 10^9$ & 2.4 & 1$\times 10^5$ & 0.02 & 0.2 & 0.2 & 0 & 173 \\ 
    M22 & 5$\times 10^9$ & 2.4 & 5$\times 10^5$ & 0.02 & 0.2 & 0.2 & 0 & 159 \\ 
    M23 & 5$\times 10^9$ & 2.4 & 1$\times 10^6$ & 0.02 & 0.2 & 0.2 & 0 & 580 \\ 
    M24 & 5$\times 10^9$ & 2.4 & 1$\times 10^6$ & 0.02 & 0.2 & 0.03 & 0 & 159 \\ 
    M25 & 5$\times 10^9$ & 2.4 & 5$\times 10^5$ & 0.02 & 0.2 & 0.03 & 0 & 159 \\ 
    M26 & 5$\times 10^9$ & 2.4 & 1$\times 10^6$ & 0.02 & 0.2 & 0.03 & 0 & 991 \\ 
    M27 & 5$\times 10^9$ & 2.4 & 1$\times 10^6$ & 0.02 & 0.2 & 0.2 & 60 & 105 \\ 
    M28 & 5$\times 10^9$ & 1.7 & 1$\times 10^6$ & 0.02 & 0.1 & 0.2 & 30 & 141 \\ 
    M29 & 5$\times 10^9$ & 2.4 & 1$\times 10^6$ & 0.02 & 0.2 & 0.2 & 0 & 78 \\ 
    M30 & 5$\times 10^9$ & 2.4 & 1$\times 10^6$ & 0.02 & 0.2 & 0.2 & 0 & 81 \\ 
    M31 & 5$\times 10^9$ & 2.4 & 1$\times 10^6$ & 0.02 & 0.2 & 0.2 & 0 & 81 \\ 
    M32 & 5$\times 10^9$ & 2.4 & 1$\times 10^6$ & 0.02 & 0.2 & 0.2 & 0 & 79 \\ 
    M33 & 5$\times 10^9$ & 2.4 & 3$\times 10^5$ & 0.02 & 0.2 & 0.2 & 0 & 108 \\ 
    M34 & 5$\times 10^9$ & 3.4 & 1$\times 10^6$ & 0.02 & 0.2 & 0.2 & 0 & 86 \\ 
    M35 & 5$\times 10^9$ & 2.4 & 3$\times 10^6$ & 0.02 & 0.2 & 0.2 & 0 & 385 \\ 

    \hline
    \end{tabular}
    \label{table:all_sims}
\end{table*}

\begin{table*}
    \centering
	\caption{Parameters for our AGB models. See the Galactic GC model table for a description of the headers.}
    \begin{tabular}{lcccccccr}
    \hline
    Model ID & $M_{\rm{dm}}$ ($\rm{M}_{\odot}$) & $\rm{R}_{galaxy}$ (kpc) & $M_{gc}$ ($\rm{M}_{\odot}$) & $\rm{R}_{gc}$ (pc)& $\rm{R}_{HI \ hole}$ (kpc)& $f_{\rm rot}$ & $\theta_{\rm gc}$ ($^{\circ}$)& $\rm{R}_{start}$ (pc) \\ 
    \hline
    M36 & 5$\times 10^9$ & 1.7 & 1$\times 10^6$ & 0.02 & 0.5 & 0.2 & 30 & 36 \\ 
    M37 & 5$\times 10^9$ & 2.4 & 1$\times 10^6$ & 0.02 & 0.5 & 0.03 & 0 & 241 \\ 
    M38 & 5$\times 10^9$ & 2.4 & 5$\times 10^6$ & 0.02 & 0.5 & 0.03 & 0 & 223 \\ 
    M39 & 5$\times 10^9$ & 2.4 & 5$\times 10^6$ & 0.02 & 0.5 & 0.2 & 0 & 223 \\ 
    M40 & 5$\times 10^9$ & 2.4 & 1$\times 10^6$ & 0.02 & 0.5 & 0.2 & 60 & 402 \\ 
    M41 & 5$\times 10^9$ & 2.4 & 1$\times 10^6$ & 0.02 & 1.0 & 0.2 & 0 & 85 \\ 
    M42 & 5$\times 10^9$ & 2.4 & 1$\times 10^6$ & 0.02 & 2.0 & 0.2 & 0 & 79 \\ 
    M43 & 5$\times 10^9$ & 2.4 & 3$\times 10^6$ & 0.02 & 1.0 & 0.2 & 0 & 383 \\  
    \hline
    \end{tabular}
    \label{table:HI_sims}
\end{table*}

\begin{table*}
    \centering
	\caption{Parameters for our LSB galaxy models. See the Galactic GC model table for a description of the headers.}
    \begin{tabular}{lcccccccr}
    \hline
    Model ID & $M_{\rm{dm}}$ ($\rm{M}_{\odot}$) & $\rm{R}_{galaxy}$ (kpc) & $M_{gc}$ ($\rm{M}_{\odot}$) & $\rm{R}_{gc}$ (pc)& $\rm{R}_{HI \ hole}$ (kpc)& $f_{\rm rot}$ & $\theta_{\rm gc}$ ($^{\circ}$)& $\rm{R}_{start}$ (pc) \\ 
    \hline
    M44 & 5$\times 10^9$ & 2.4 & 3$\times 10^5$ & 0.02 & 0.2 & 0.2 & 0 & 1907 \\ 
    M45 & 5$\times 10^9$ & 4.9 & 1$\times 10^6$ & 0.02 & 0.2 & 0.2 & 0 & 196 \\ 
    M46 & 5$\times 10^9$ & 2.4 & 1$\times 10^5$ & 0.02 & 0.2 & 0.2 & 0 & 1941 \\ 
    M47 & 5$\times 10^9$ & 2.4 & 1$\times 10^6$ & 0.02 & 0.2 & 0.2 & 0 & 1951 \\ 
    \hline
    \end{tabular}
    \label{table:LSB_sims}
\end{table*}


\bsp	
\label{lastpage}
\end{document}